\documentclass[aps,pra,twocolumn,showpacs,amsmath,amssymb,floatfix,superscriptaddress,notitlepage,nofootinbib]{revtex4-1}

\usepackage{color}
\usepackage{bbm}
\usepackage{graphicx}% Include figure files
\usepackage{dcolumn}% Align table columns on decimal point
\usepackage[utf8]{inputenc}
\usepackage[normalem]{ulem}

\usepackage{graphicx}
\usepackage{epstopdf}
\usepackage{amsthm}
\usepackage{amsmath}
\usepackage{empheq}
\usepackage{bbm}
\usepackage{braket}
\usepackage{amssymb}
\usepackage[usenames,dvipsnames]{xcolor}
\usepackage{cases}
\usepackage{latexsym}
\usepackage[colorlinks=true,citecolor=Cerulean,linkcolor=RubineRed,urlcolor=Cerulean]{hyperref}

\graphicspath{ {images/}{images/figure1/}{images/figure2/}{images/figure3/}{images/figure4/}{images/figure5/} }

\newcommand{\ketbra}[2]{|#1\rangle\langle #2|}

\newcommand{\tr}[1]{\operatorname{\textnormal{Tr}}\left[ {#1} \right]}  %Trace
\newcommand{\corr}{\mathcal{C}}  

\usepackage{lipsum}
\usepackage{graphicx}

\begin{document}

\title{
Relaxation of non-integrable systems and correlation functions
}

\date{\today}

\author{Jonathon Riddell}
\email{riddeljp@mcmaster.ca}
\affiliation{Department of Physics \& Astronomy, McMaster University
	1280 Main St.  W., Hamilton ON L8S 4M1, Canada.}
\affiliation{Perimeter Institute for Theoretical Physics, Waterloo, ON N2L 2Y5, Canada}

\author{Luis Pedro Garc\'ia-Pintos}
\email{lpgp@umd.edu}
\affiliation{Joint Center for Quantum Information and Computer Science and Joint Quantum Institute, NIST/University of Maryland, College Park, Maryland 20742, USA}

\author{\'Alvaro M. Alhambra}
\email{alvaro.alhambra@mpq.mpg.de}
\affiliation{Max-Planck-Institut für Quantenoptik, Hans-Kopfermann-Straße 1, D-85748 Garching, Germany}

\begin{abstract}
We investigate early-time equilibration rates of observables in closed many-body quantum systems and compare them to those of two correlation functions, first introduced by Kubo and Srednicki. We explore whether these different rates coincide at a universal value that sets the timescales of processes at a finite energy density. We find evidence for this coincidence when the initial conditions are sufficiently generic, or typical. We quantify this with the effective dimension of the state and with a state-observable effective dimension, which estimate the number of energy levels that participate in the dynamics. 
Our findings are confirmed by proving that these different timescales coincide for dynamics generated by Haar-random Hamiltonians. This also allows to quantitatively understand the scope of previous theoretical results on equilibration timescales and on random matrix formalisms.  
We approach this problem with exact, full spectrum diagonalization. The numerics are carried out in a non-integrable Heisenberg-like Hamiltonian, and the dynamics are investigated for several pairs of observables and states. 
\end{abstract}

\maketitle

The current rate of development of quantum technologies means that experiments on quantum many body systems away from equilibrium are within reach. One of the more easily realizable mechanisms in this context is that of a \emph{quantum quench} \cite{mitra2018quantum}. In it, one prepares a simple initial state $\ket{\Psi}$ of a lattice system, such as a low energy eigenstate of Hamiltonian $H_0$, and the Hamiltonian is suddenly switched to $H$ for which $\ket{\Psi}$ is no longer an eigenstate. This drives the system far from equilibrium, and the subsequent dynamics can be traced through the expectation values of observables $A$,
\begin{equation}\label{eq:expvalue}
    \langle A (t) \rangle \equiv \bra{\Psi} e^{-iHt}A e^{iHt} \ket{\Psi},
\end{equation}
where $t$ is the time elapsed after the quench. 

The experimental relevance of this setting has triggered a large amount of theoretical work, aimed at describing the complex dynamics of a wide variety of models \cite{gogolin2016equilibration,d2016quantum}. 
One of the most prominent features of these quantum dynamics is that physically relevant observables often thermalize, in the sense that there is some time $T$ such that $
    \langle A(t) \rangle \sim \langle A \rangle_\beta$ for $t \ge T$,
where $\langle A \rangle_\beta \nobreak=\nobreak \tr{\frac{e^{-\beta H}}{Z}A}$ is the expectation value of a Gibbs ensemble with average energy $\langle \Psi \vert H \vert \Psi \rangle$. It is by now established that thermalization occurs generically as a consequence of the \emph{eigenstate thermalization hypothesis} (ETH) \cite{Deutsch91,srednicki1994chaos}.
The ETH implies that 
energy eigenstates within a microcanonical window have thermal expectation values, from which it follows that $\overline{\langle A(t)\rangle} \nobreak\rightarrow\nobreak \langle A \rangle_\beta$, where $\overline{\langle ...\rangle}$ denotes the long-time average. 

Less is known about how fast systems thermalize and about how the ETH affects the approach to thermal equilibrium (also called relaxation or equilibration).  
It has been argued that the ETH is behind the fast relaxation
to steady state values~\cite{Reimann_2015,reimann2016typical,ShortFarrelly11,Garcia-PintosPRX2017,Wilming_2018,richter2018impact,heveling2020compelling,heveling2020comment,Dabelow2020}. It has also been shown to play a role in fluctuation-dissipation relations~\cite{Srednicki99,Khatami13,d2016quantum,Nation_2019,Noh_2020}, %the presence of 
certain kinds of transport~\cite{d2016quantum,bertini2020finitetemperature,schonle2020eigenstate,Dymarsky2018bound}, and in the appearance of random matrix-like phenomena~\cite{d2016quantum,Mondaini17,Dymarsky2018bound,Richter_2020,brenes2021outoftimeorder}. However, our theoretical understanding of these  processes, their timescales, and of how the ETH exactly influences them remains far from complete.

Here, we investigate the relaxation timescales of $A$ by numerically analyzing the early-time decay of the expectation value $\langle A (t)\rangle$ for various observables and states. We focus on a non-integrable Heisenberg model that we study via exact diagonalization.    
For the cases studied, we observe numerically that at early times the expectation value decays as
\begin{equation}
\label{eq:initialdecay}
    \langle A(t) \rangle \simeq \langle A(0) \rangle e^{-  \frac{\sigma_A^2 t^2}{2}} \simeq \langle A(0) \rangle \left(1-  \frac{\sigma_A^2 t^2}{2} \right) ,
\end{equation}
for some constant $\sigma_A>0$. This sets the initial \emph{relaxation rate}, which dominates until later phenomena, such as hydrodynamic tails~\cite{lux2014hydrodynamic,Blake_2016,Dymarsky2018bound}, become relevant. We thus identify $
\sigma_A$ as the object of study of previous works on the timescales of equilibration~ \cite{ShortFarrelly11,Garcia-PintosPRX2017,de2018equilibration,Wilming17,Wilming_2018}.

We study the relaxation rate $\sigma_A$, and relate it to those of two correlation functions: $\corr(t)$ describing the long-time out of equilibrium fluctuations, introduced by Srednicki~\cite{Srednicki99}, and the Kubo function $\corr_{\text{Kubo}}(t)$ describing the dissipation of perturbations near thermal equilibrium. We also calculate these rates analytically in a random matrix theory model, for which we show that they coincide up to a $d^{-1}$ error, with $d$ the Hilbert space dimension. Inspired by this result and other insights from random matrix theory frameworks~\cite{reimann2016typical,nation2018quantum}, we theorize, and numerically analyze, that in ``generic'' situations, the decay rate of all these quantities closely matches. This sets a universal timescale, which may only depend on few parameters such as the temperature~\cite{pappalardi2021quantum}.

The coincidence of correlation functions introduced by Srednicki and Kubo has been previously referred to as a ``fluctuation-dissipation theorem"~\cite{Khatami13,d2016quantum}. With our analysis, we go beyond this connection by exploring how this fluctuation-dissipation relation can already arise during the initial relaxation process of the system. This is also supported by previous results~\cite{richter2018impact,richter2019relation} which found other initial conditions under which thermal correlation functions coincide with the post-quench evolution of observables.

To study the relation between the different decay rates, we numerically determine $\sigma_A$ for different initial conditions and system sizes, up to $L=24$, and compare it to the rates $\sigma_G$ and $\sigma_{K}$ which characterize the dynamics of the Srednicki and Kubo correlation functions, respectively. We find that they are of the same order of magnitude in all cases studied, and they converge to the same value for at least one of the observables and initial state considered. 
We also observe that they are closer the more generic or ``typical" the initial conditons are, which we quantify with the so-called \emph{effective dimension}~\cite{Reimann08,short2011equilibration} of the initial state, and with a modified version of it that also depends on the observable measured.

With these results, we illustrate with exact numerics previous theoretical arguments regarding timescales of relaxation \cite{Srednicki99,de2018equilibration,Wilming17}, narrowing down their regime of applicability.
We conclude that the accuracy of many existing theoretical predictions, (in particular, their ability to predict the relaxation timescale accurately) crucially depends on the typicality of the initial conditions, which appears to be challenging to quantify rigorously. We propose the aforementioned two quantities as figures of merit of this typicality, and we conclude that their value also relates to the validity of random matrix theory models studied in the literature~\cite{Deutsch-ETH,Popescu_2006,Reimann_2015,reimann2016typical,Reimann_2019,nation2018off,Nation_2019}. 

The paper is structured as follows. We define the model, observables, and states considered in our simulations in Sec.~\ref{sec:model}, and we study the decay rates of the observables in Sec.~\ref{sec:ratefit}.
In Sec.~\ref{sec:corr}, we define the correlation function of Srednicki and in Sec.~\ref{sec:kubo} that of Kubo. We compare the different decay rates in Sec.~\ref{sec:cete}.
In Sec.~\ref{sec:RMT}, we show that all timescales coincide for Haar random Hamiltonians of large dimensional systems. %, which relies on an analytic proof of the ``annealed approximation'' in random matrix theory.}
In Sec.~\ref{sec:effdim}, we further investigate the conditions under which $\langle A(t) \rangle$, $\corr(t)$, and $\corr_{\text{Kubo}}(t)$ have similar decay rates.
We end with some remarks and open questions in Sec.~\ref{sec:conclusion}.

 %%%%%%%%%%%%%%%%
 %%%%%%%%%%%%%%%
 
  \section{Non-integrable model: Heisenberg chain}
 \label{sec:model}
 
 We provide numerical results using a Heisenberg chain with next-nearest neighbour interactions
 \begin{align}
	{H} =  &\sum_{j=1}^L J_1 \left(  {S}_j^+ S_{j+1}^- + \text{h.c}\right) + \gamma_1  \, {S}_j^Z    {S}_{j+1}^Z  \nonumber \\ &+  J_2 \left(  {S}_j^+  {S}_{j+2}^- + \text{h.c}\right)+ \gamma_2   {S}_j^Z  {S}_{j+2}^Z, \label{eq:hamiltonian}
\end{align}
where $S_j^Z = \ket{\uparrow}\!\bra{\uparrow} - \ket{\downarrow}\!\bra{\downarrow}$ is the Pauli operator along $Z$ for spin $j$ and $S_j^+ = \ket{\uparrow}\!\bra{\downarrow}$. We characterize this model by the coefficient vector $ (J_1,\gamma_1,J_2,\gamma_2)$. It is non-integrable and obeys the ETH for generic parameters. However, it is quasi-free for $ (J_1,\gamma_1,J_2,\gamma_2) = (J_1,0,0,0)$ and Bethe-ansatz solvable when $ (J_1,\gamma_1,J_2,\gamma_2) = (J_1,\gamma_1,0,0)$ \cite{Santos2010,Cazalilla2011,coleman_2015}.  Unless otherwise stated, we take 
$(J_1,\gamma_1,J_2,\gamma_2) \nobreak = \nobreak (-1,1,-0.2,0.5)$ for the figures.

The Hamiltonian in Eq.~\eqref{eq:hamiltonian} has a number of symmetries which allows us to block-diagonalize it.  In particular, it preserves total magnetization in the $ {m}_z \nobreak =  \nobreak \sum_{j=1}^L {S}_j^Z$ direction and is translation invariant. In our numerics, we choose initial states with $\langle  {m}_z\rangle  \nobreak = \nobreak 0$, which allows to further exploit the $Z_2$ spin flip symmetry. Our initial states are also chosen such that they have support in the $k=\{0,\pi\}$ translation sectors. This allows us to access exact dynamics from the maximally symmetric blocks of the Hamiltonian using the spatial reflection symmetry. 

We investigate two observables of interest, 
\begin{align}
\label{eq:observ1}   &A_1 \coloneqq S_1^Z, \\
  \label{eq:observ2}  &A_2 \coloneqq \frac{1}{L} \sum_{j=1}^L    {S}_j^Z  {S}_{j+1}^Z .
\end{align}
Observable $A_1$ has support on a single site and is therefore not translation invariant. This requires eigenstates from different symmetry sectors of the Hamiltonian to contribute to its dynamics. This is fundamentally different from the translation-invariant observable $A_2$, for which dynamics are generated only between eigenstates in the same symmetry sector. 

We study three initial states that, as we see below, showcase rapid and slow examples of the relaxation process when probed by the observables introduced above.
Let, 
\begin{align} \label{eq:state1}
    |\psi\rangle & \coloneqq |\uparrow\downarrow\uparrow\downarrow\dots ..\rangle, 
    \\ \label{eq:state2}
 |\psi'\rangle &\coloneqq \frac{1}{\sqrt{2}}\left( |\uparrow\downarrow\uparrow\downarrow\dots ..\rangle+|\downarrow \uparrow\downarrow\uparrow\dots ..\rangle\right), 
\\
 \label{eq:state3}
        |\phi \rangle &\coloneqq \frac{1}{\sqrt{L}}\sum_{r=0}^{L-1} \hat{T}^r|\uparrow\uparrow \dots \uparrow \downarrow \dots \downarrow \downarrow \rangle, 
\end{align}
where $\hat{T}$ is the translation operator shifting the state of each spin by one lattice site.

One expects the N\'eel-type state $|\psi\rangle$ to rapidly approach equilibrium due to every $\uparrow$ being neighboured by two $\downarrow$ terms. Also, $|\psi'\rangle$ being a superposition of two such states, we expect it to behave similarly, although its ``cat state" structure may slightly change the physics. 

Meanwhile, 
$|\phi\rangle$ is a translation invariant state where the generating  state $|\uparrow\uparrow \dots \uparrow \downarrow \dots \downarrow \downarrow \rangle$ %on the other hand 
only has two points that will initially admit spin flip dynamics. Due to this property, we expect it to exhibit slower equilibration that is more likely to feature hydrodynamic tails associated with transport. 

In summary, we focus on the observable-state pairs 
\begin{subequations}
\label{eq:obs-state}
\begin{align}
\label{eq:at}
    \langle A_1(t) \rangle_{\ket \psi}  &\coloneqq \langle \psi |  {A}_1 (t) |\psi  \rangle, \\ 
    \label{eq:bt}
    \langle A_2 (t) \rangle_{\ket \phi}     &\coloneqq \langle \phi  |  {A}_2 (t) |\phi  \rangle, \\
    \label{eq:gt}
    \langle A_2  (t) \rangle_{\ket{\psi'}}     &\coloneqq  \langle \psi' \vert  {A}_2 (t)  \vert \psi' \rangle.
\end{align}
\end{subequations}

\section{Initial decay rate}
\label{sec:ratefit}

We observe that the initial decay of expectations after a quench takes the form
\begin{align}
\label{eq:initialdecay2}
    \langle A(t) \rangle &\simeq \left(\langle A(0) \rangle - \langle A(\infty) \rangle  \right) e^{-  \frac{\sigma_A^2 t^2}{2}} +\langle A(\infty) \rangle
    \\ & \simeq \left(\langle A(0) \rangle - \langle A(\infty) \rangle  \right) \left(1-  \frac{\sigma_A^2 t^2}{2} \right) +\langle A(\infty) \rangle, \nonumber
\end{align}
where $\langle A(\infty) \rangle$ represents a late time ``equilibrated" value that for simplicity we will take to be $\langle A(\infty) \rangle=0$ throughout. 

A Taylor expansion to second order shows that
\begin{align}
\label{eq:sigmaA}
\sigma_A^2 &= - \frac{1}{\langle A(0) \rangle} \frac{d^2 \langle A(t) \rangle}{dt^2} \Big \vert_{t=0}  = -\frac{\langle [H,[H,A]] \rangle}{\langle A(0) \rangle}
\nonumber \\ &= \frac{ \sum_{j,k} c_j c^*_k A_{jk} (E_j-E_k)^2}{\sum_{j,k} c_j c^*_k A_{jk}},
\end{align}
where in the last line we expand in the energy eigenbasis $\{ \ket{E_j} \}$ such that $H=\sum_j E_j \ket{E_j} \langle E_j \vert$, the initial state is $\ket{\Psi}=\sum_j c_j \ket{E_j}$ and $A_{jk}=\bra{E_j} A \ket{E_k}$. This second derivative has previously appeared in the analysis of general equilibration timescales \cite{Garcia-PintosPRX2017}.

\begin{figure}
 \centering
            \centering
            \includegraphics[width=0.5\textwidth]{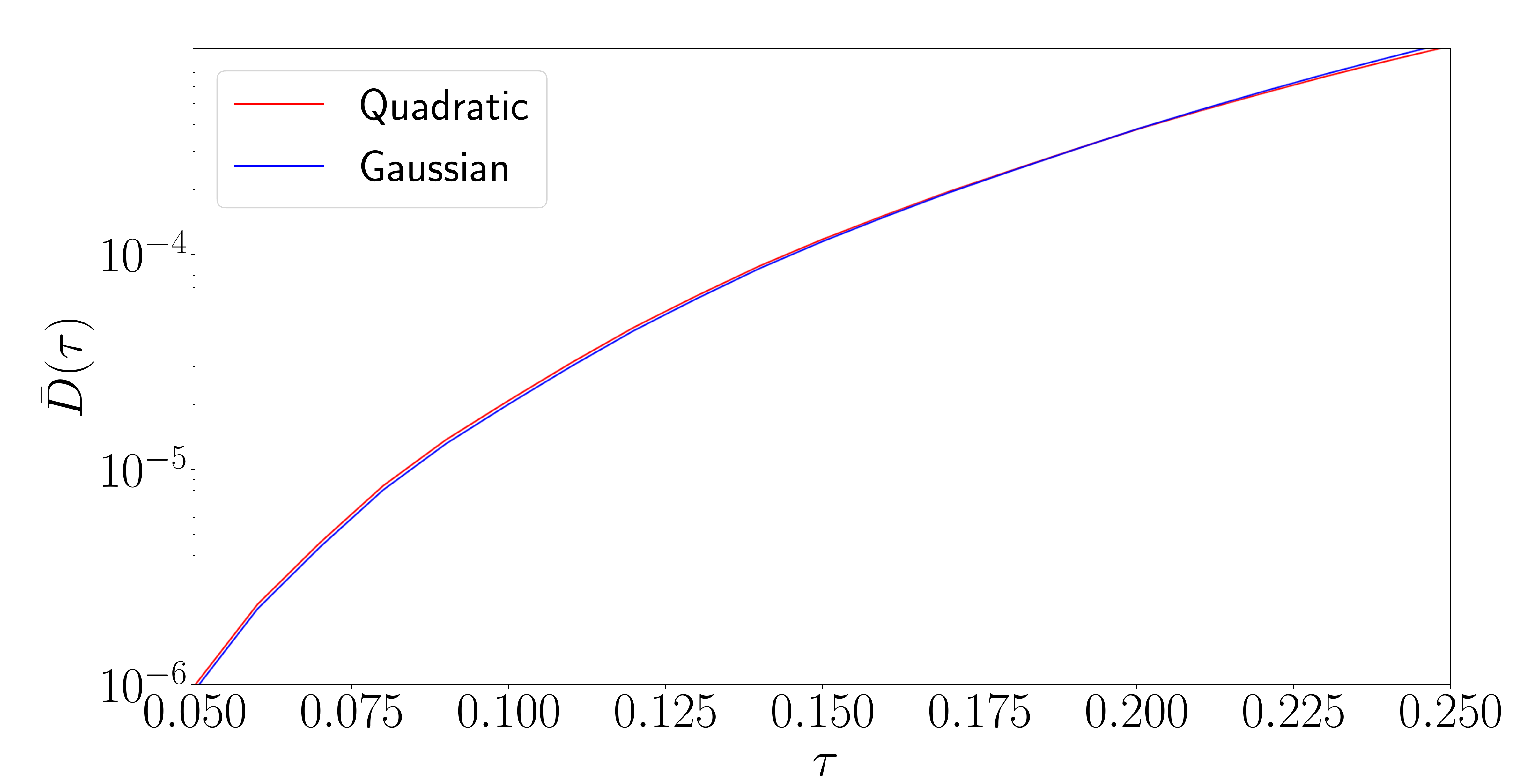}
            \caption[Network2]%
            {{\small Early time fit of $\langle A_1(t) \rangle_{\ket \psi}$ as defined in Eq.~\eqref{eq:at} with respect to a quadratic and a Gaussian. We define a discrete approximation of the functions to be compared at the times $t \in \{ n \Delta t  | n\in \mathcal{N} \land  n \Delta t \leq \tau  \}$. We take $ \Bar{D}(\tau)  = ||\langle \overrightarrow{ A_1}(t) \rangle_{\ket \psi}-\overrightarrow{f}(t)||_1/ || \langle A_1(t) \rangle_{\ket \psi}||_1 $ where $\overrightarrow{g}(t)$ is the vector with components $g(n \Delta t)$, and $f(t)$ is the fitted function on the discrete interval. Here $\Delta t = 0.01$ and the fits are performed for $L = 24$ dynamics. 
              }}    
           \label{fig:fit}
    \end{figure}

Since we focus on an early-time regime, both the Gaussian and the quadratic functions give good approximations, with the Gaussian being slightly more accurate in most cases. 
The accuracy of the approximation is shown in Fig.~\ref{fig:fit} for the observable-state pair in Eq.~\eqref{eq:at} (the other two show similar behaviour). Here, we see that the Gaussian does a marginally better job at approximating the decay in the dynamical region of interest and both functions get progressively worse at approximating the dynamics as $t$ increases. This differs from the exponential decay $\sim e^{-\sigma t}$ associated with Fermi's golden rule found in some regimes~\cite{Bartsch_2008,Mallayya_2019,heveling2020compelling}. In the cases studied here, a leading linear term is already ruled out from the fact that for our initial states and observables it holds that $[A,\ket{\Psi}\!\bra{\Psi}]=0$.

\begin{figure*}%[ht]
  \centering
        \includegraphics[trim=00 00 00 00,width=0.325\textwidth]{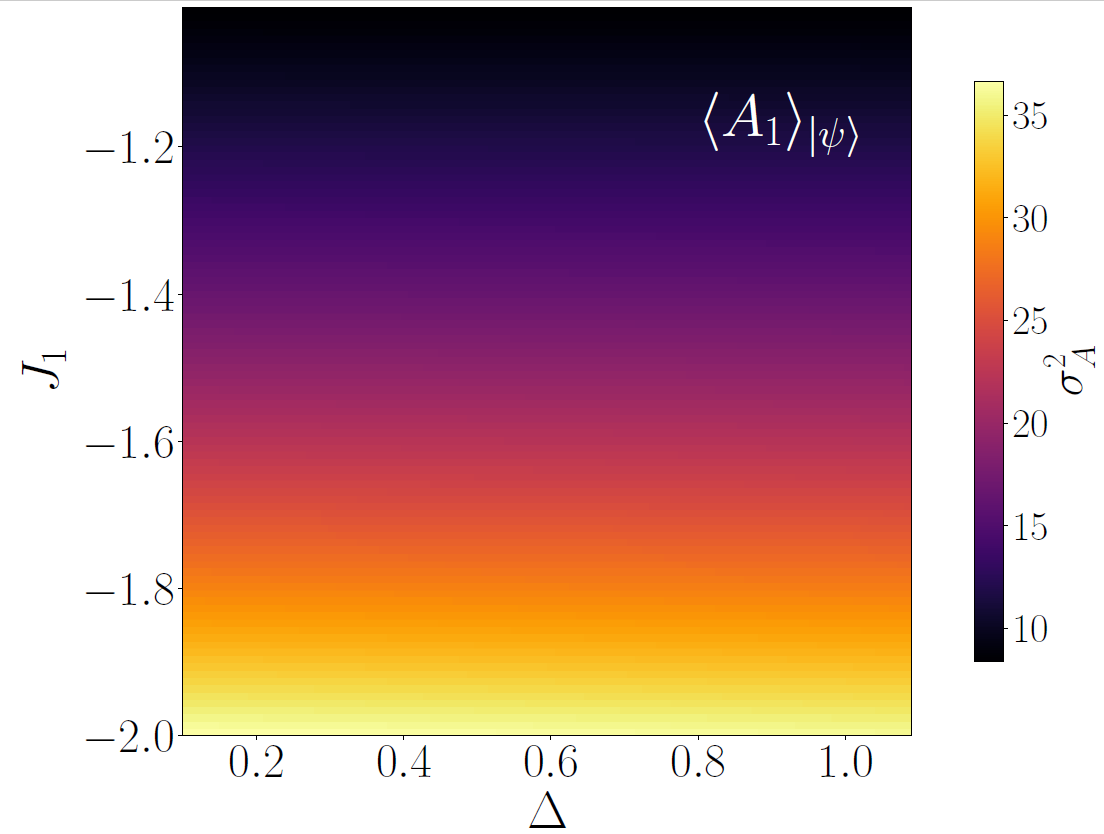}  
         \includegraphics[trim=00 00 00 00,width=0.325\textwidth]{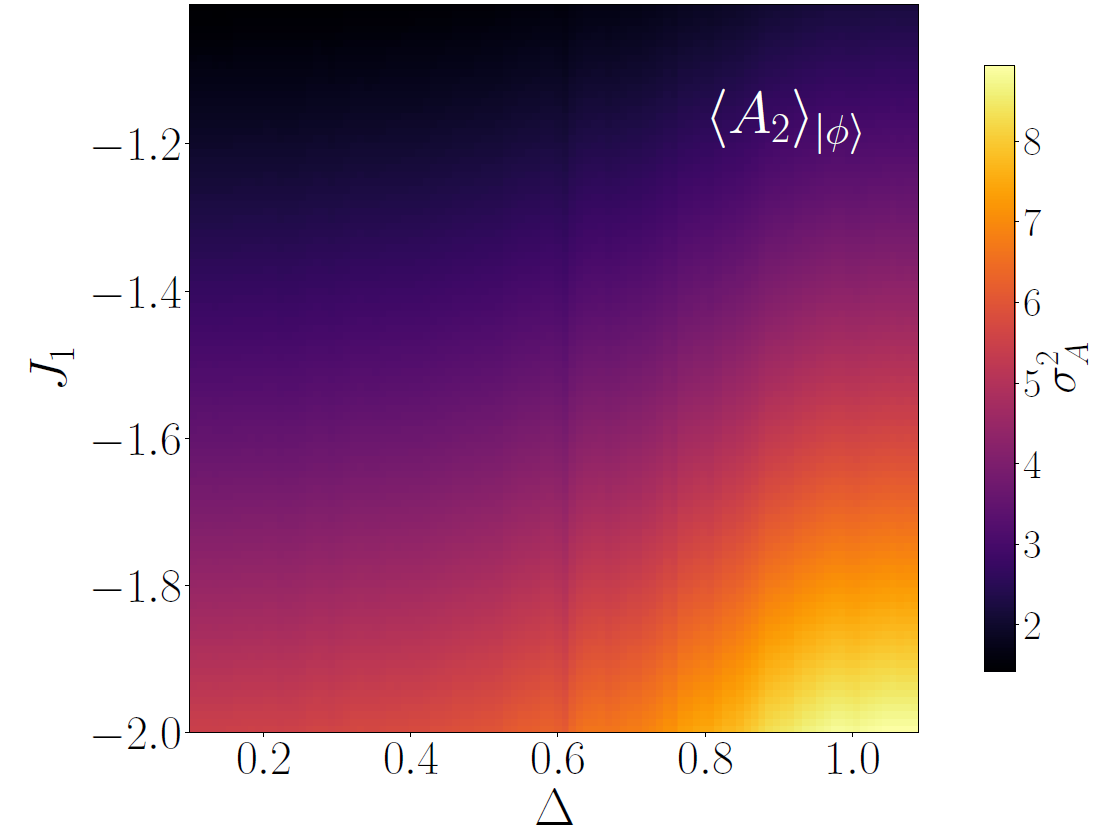}  
         \includegraphics[trim=00 00 00 00,width=0.325\textwidth]{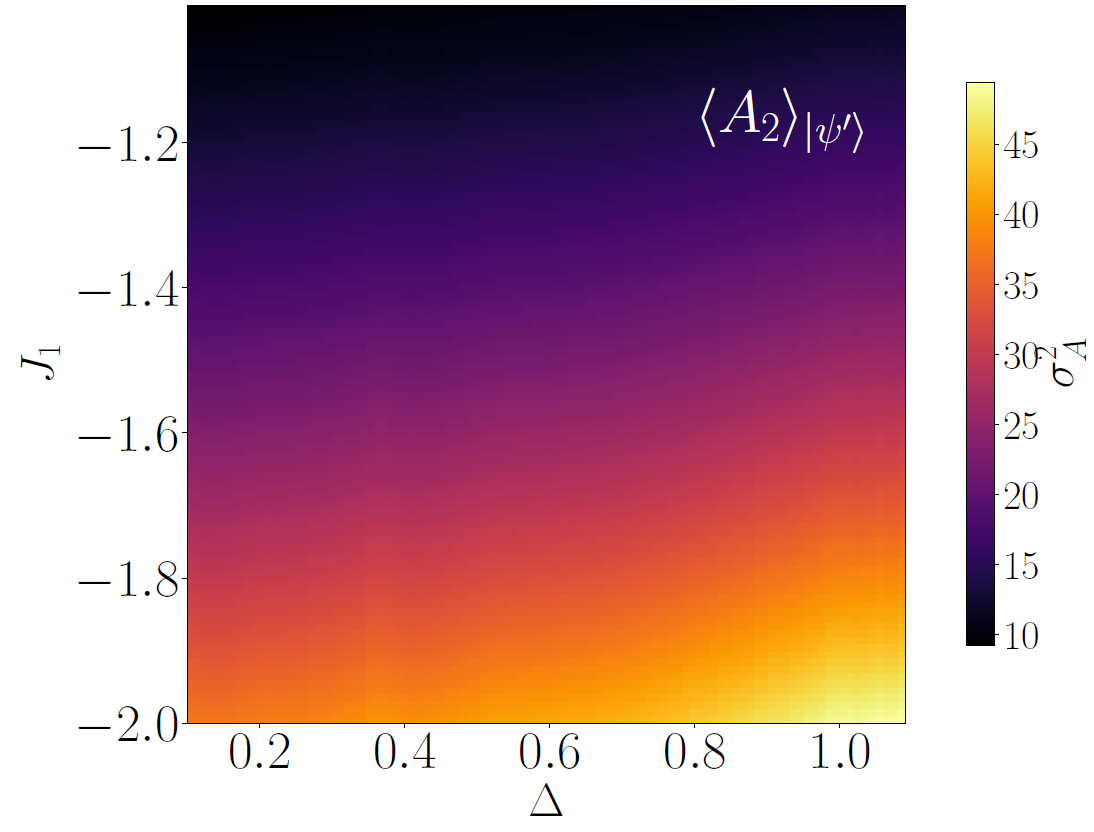} 
      \\
       \includegraphics[trim=00 00 00 00,width=0.325\textwidth]{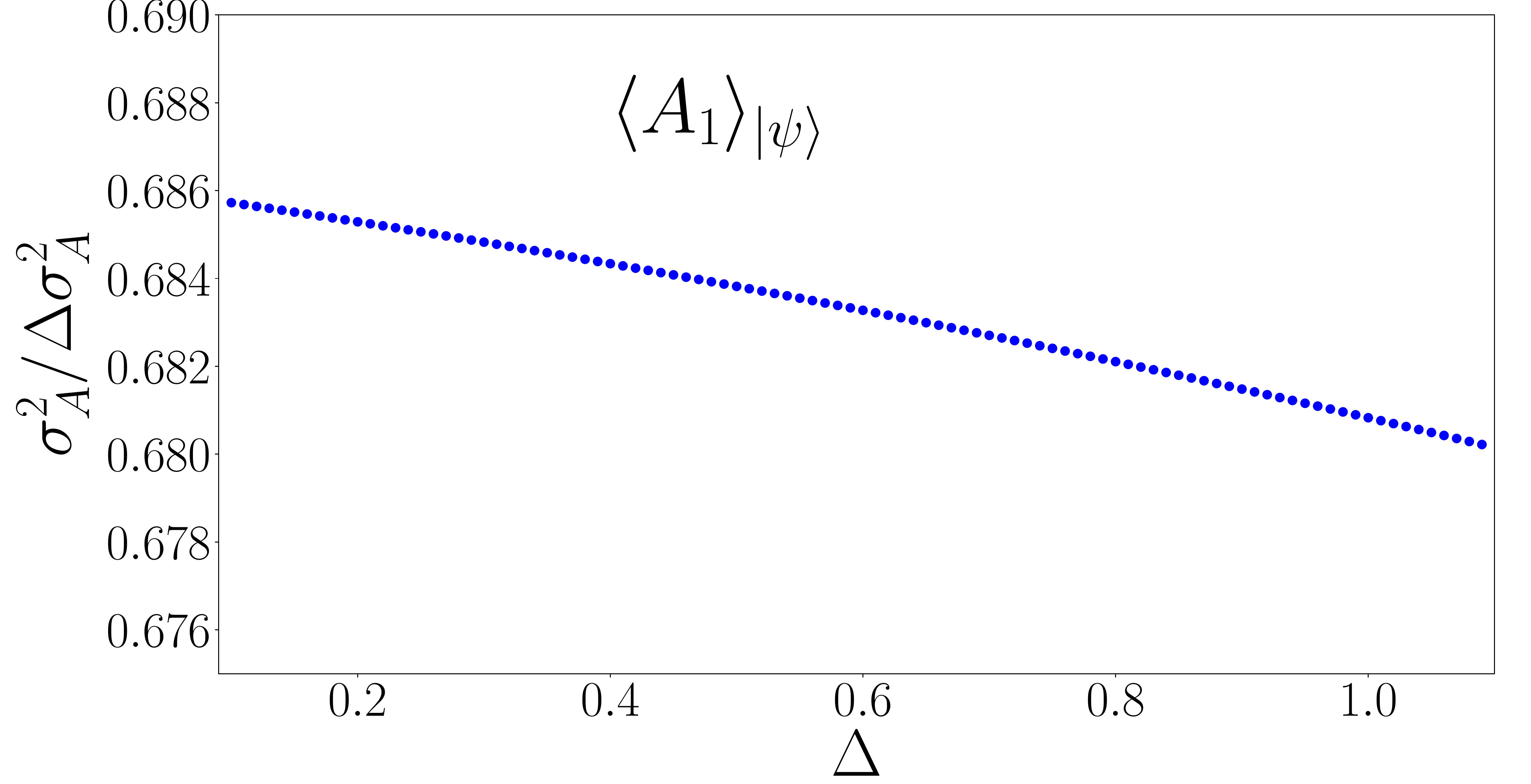}  
         \includegraphics[trim=00 00 00 00,width=0.325\textwidth]{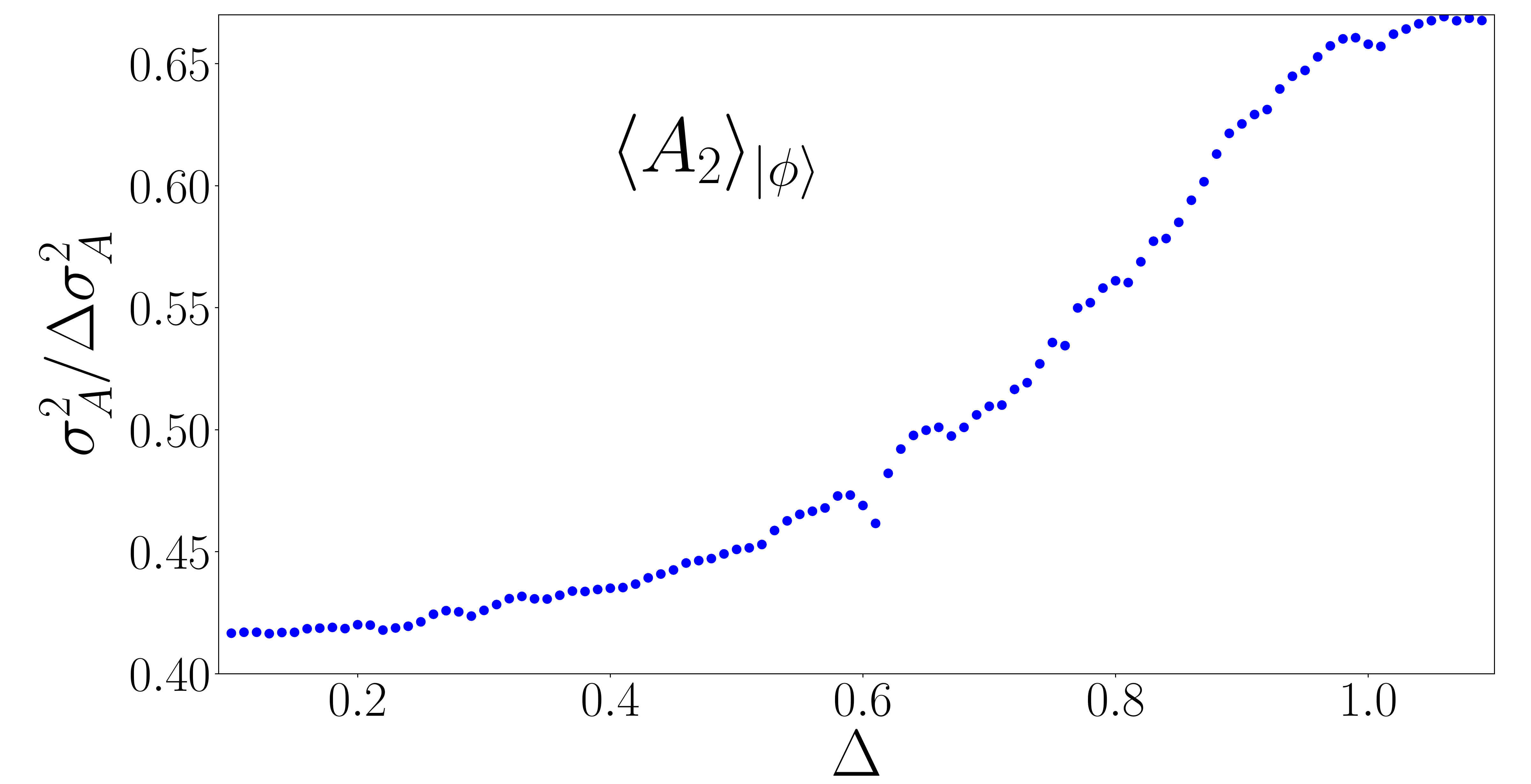}  
         \includegraphics[trim=00 00 00 00,width=0.325\textwidth]{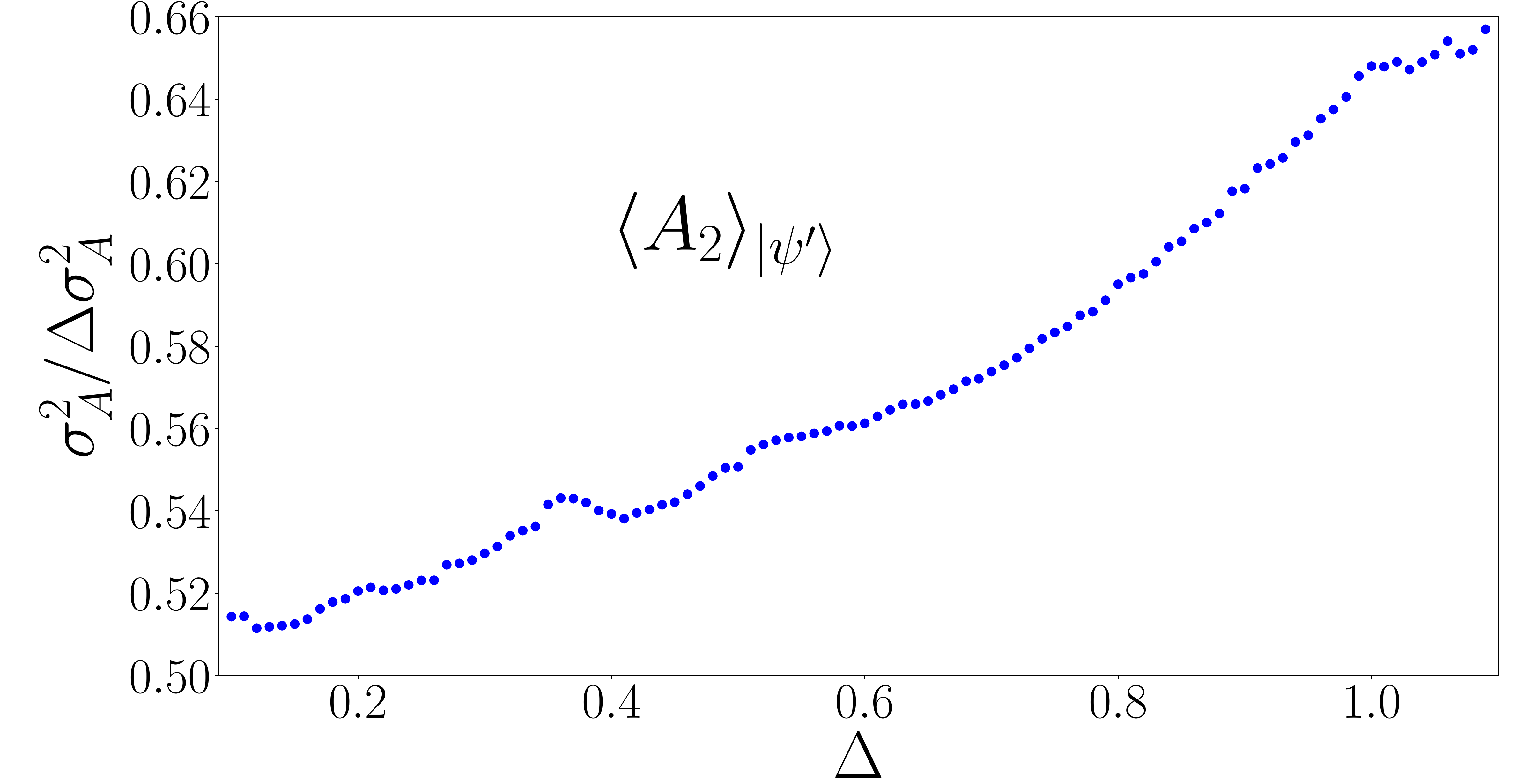} 
\caption{
\label{fig:changeKappa} 
{\bf 
 }
The plots use the parameters of  $(J_1,\gamma_1,J_2,\gamma_2) =( J_1,\frac{J_1\Delta}{2}, J_2,\frac{J_2\Delta}{2}   )$ with $ J_2 = \frac{1}{2.7} J_1$ fixed. All data shown here has a system size of $L = 18$. (First row) Heat map of $\sigma_A^2$ as a function of $J_1$ and $\Delta$ for our three observable/state pairs. (Second row) We define $\Delta \sigma_A^2 = \max \sigma_A^2 - \min \sigma_A^2$, where the maximum and minimum are extracted from the data of the first row. To illustrate the individual variation of $\sigma_A^2$ with respect to $\Delta$ we cut the data from the first row by fixing $J_1 = -1.8$. Normalizing the $\sigma_A^2$ by dividing out $\Delta \sigma_A^2$ indicates how much change in the value is due to changes in $\Delta$ compared to $J_1$.
}
\end{figure*}

We first investigate the dependence of the initial decay rate $\sigma_A$ with respect to various parameters of the Hamiltonian in Eq.~\eqref{eq:hamiltonian}. To produce a systematic picture of the dependence of $\sigma_A$ on the parameters of the Hamiltonian, we take $(J_1,\gamma_1,J_2,\gamma_2) =( J_1,\frac{J_1\Delta}{2}, J_2,\frac{J_2\Delta}{2}   )$.  We further fix the relation $ J_2 = \frac{1}{2.7} J_1$. The results are shown in Fig.~\ref{fig:changeKappa} for $L=18$. We vary $J_1 \in [-2,-1]$ and $\Delta \in [0.1,1.1]$. In this regime we see that  the $\sigma_A^2$ of $\langle A_1(t) \rangle_{\ket \psi} $ is largely independent of $\Delta$ in the tested regime. On the contrary, the decay rate $\sigma_A$ of $\langle A_2 (t) \rangle_{\ket \phi}$ varies quite strongly with respect to $\Delta$, most likely due to its construction hindering spin flip dynamics early. Our third example sits in-between these two. While the timescales associated with $\langle A_2  (t) \rangle_{\ket{\psi'}}  $ vary weakly with respect to $\Delta$, the effect is non-negligible. 
Unsurprisingly, the timescales are much more sensitive to $J_1$ since it directly controls the magnitude of interactions which do not commute with the observables studied. 
As we will see, we can associate this lack of dependence on certain Hamiltonian parameters with the coincidence of timescales that we explore. 

%%%%%%%%%%%%%%%%
%%%%%%%%%%%%%%%%%

\section{Fluctuations: Srednicki's correlation function}\label{sec:corr}

We now consider the correlation function $\corr(t)$, first defined in \cite{Srednicki99}, which quantifies the ``correlations in time" of $\langle A(t) \rangle$, as \footnote{Due to ETH, and the assumption that $\langle A(\infty) \rangle=0$, one could equivalently write these sums including the terms with $j=k$.}

 \begin{align}
 \label{eq:Corr}
     \corr(t)= \frac{\overline{\langle A (t+t')\rangle \langle A (t')\rangle}}{\overline{\langle A (t')\rangle^2}},
 \end{align}
where $\overline{X(t')}\equiv \lim_{T \rightarrow \infty} \int_0^T \frac{\text{d}t'}{T}X(t')$. It can also be written as
\begin{equation}\label{eq:coft}
    \corr(t)=\frac{\sum_{j\neq k}\vert c_j\vert^2 \vert c_k\vert^2 \vert A_{jk}\vert^2 e^{-i t (E_j-E_k)} }{\sum_{j \neq k}\vert c_j\vert^2 \vert c_k\vert^2 \vert A_{jk}\vert^2}.
\end{equation}

The ETH and properties of the observable and the state imply that, for short times, this function decays as $\corr(t) \simeq e^{- \frac{\sigma_G^2 t^2}{2}} \simeq 1- \frac{\sigma_G^2 t^2}{2}$ (since it is at early times, the quadratic function is also a good approximation). This is discussed in more detail in Appendix \ref{app:gaussian}. Moreover, the initial decay rate is given by
\begin{align}
\label{eq:sigmag}
    \sigma_G^2 &= \frac{\sum_{j \neq k} \vert c_j \vert^2 \vert c_k\vert ^2 \vert A_{jk} \vert^2 (E_k-E_j)^2}{\sum_{j \neq k} \vert c_j \vert^2 \vert c_k\vert ^2 \vert A_{j  k} \vert^2 } \nonumber
  \\ &=\frac{\tr{\mathcal{D}(\ketbra{\Phi}{\Phi})[A,H]\mathcal{D}(\ketbra{\Phi}{\Phi})[H,A]}}{\tr{(\mathcal{D}(\ketbra{\Phi}{\Phi})A)^2}},
\end{align}
where $\mathcal{D}(\ketbra{\Phi}{\Phi}) \coloneqq \sum_k \vert c_k \vert^2 \ketbra{E_k}{E_k}$ is the so-called \emph{diagonal ensemble}. Note that $\langle A(\infty) \rangle=\tr{\mathcal{D}(\ketbra{\Phi}{\Phi})A}=0$. This can be seen by expanding Eq.~\eqref{eq:coft} to second order at short times. With this expression, $\sigma_G$ could be calculated by more efficient methods than exact diagonalization, such as tensor networks~\cite{_akan_2021}.

It will be often the case that $\sigma_G$ characterizes relaxation timescales beyond the short-time behaviour of $\corr(t)$. We prove in Appendix~\ref{app:qslC} that the rate at which the correlation function changes is upper bounded by $ \left| \frac{d \corr(t)}{dt} \right| \leq \sigma_G$ at all times -- a form of quantum speed limit on $\corr(t)$. The main results of~\cite{alhambra2020time} also apply to this function. They show that $\sigma_G$ characterizes not only the initial rate but also the timescales for $\corr(t)$ to equilibrate to the steady state value in generic situations (see Appendix~\ref{app:equilibrationC}).

Unlike the form of $\sigma_A$ in Eq.~\eqref{eq:sigmaA}, $\sigma_G$ in Eq.~\eqref{eq:sigmag} implies that the characteristic rate $\sigma_G$ is independent of the terms in the Hamiltonian that commute with $A$. In our examples, both observables commute with the term $\gamma_1 S_j^Z S_{j+1}^Z$ and as such $
\sigma_G$ is independent of $\gamma_1$. We can now compare this conclusion to that of Fig.~\ref{fig:changeKappa}, where we see that $\sigma_A$ is mostly independent of $\gamma_1$ only for $\langle A_1(t) \rangle_{\ket \psi} $. This already hints to situations where we expect $\sigma_A \sim \sigma_G$.

In~\cite{Srednicki99}, it was argued that $\corr(t)$ exactly models the decay of $\langle A(t) \rangle$ after possible large out-of-equilibrium fluctuations that happen after the system has thermalized. 
We have not been able to numerically verify this claim, since it appears that the potential fluctuations that this correlation function models happen at late times (likely scaling quickly with system size). In any case, from the definition in Eq.~\eqref{eq:Corr}, we can establish that $\corr(t)$ quantifies the correlations in time of the fluctuations of $\langle A(t) \rangle$.

Notice that this function depends on the same initial conditions as $\langle A(t) \rangle$ (with the difference that it does not depend on the phases of $c_j c_k^* A_{jk}$). We may then expect what is one of our main points: that in some ``generic" cases, e.g. in which the dependence with the phases is unimportant, its decay rate may be close to $\sigma_A$ in Eq.~\eqref{eq:initialdecay}. This is in fact the conclusion reached by slightly different arguments of previous works on relaxation timescales \cite{de2018equilibration}. These suggests that, given ETH, the relevant decay timescale in which $\langle A(t) \rangle \nobreak\rightarrow\nobreak \langle A \rangle_\beta$ is close to precisely $\sigma_G$ (see Eq.~(15) of~\cite{de2018equilibration}).

We now explain why this can be expected. As discussed in e.g.~\cite{de2018equilibration}, 
%the problem of determining it 
the initial relaxation rate can be understood as follows: for many-body systems, the expectation value of an observable
\begin{equation}\label{eq:aoftE}
    \langle A (t) \rangle = \sum_{jk} c_j c^*_k A_{jk} e^{-i t (E_j-E_k)},
\end{equation}
at time time $t>0$ is a sum over a dense set of complex numbers oscillating at different frequencies in the complex plane. If these complex numbers are spread evenly enough among the plane, they will typically collectively cancel. This will cause the oscillating part of $ \langle A (t) \rangle$ to average to zero, leading to equilibration to a steady value. This will happen quickly in general, provided that there are not too many spurious correlations among the coefficients $c_j c^*_k A_{jk}$ and the energy gaps, that might make the spread uneven. We thus identify this spread and lack of correlations between the complex coefficients with the concept of ``typicality". A similar picture is also provided in~\cite{Wilming17}. 
  
The conclusion stemming from this is that the relaxation or ``dephasing" time is controlled by the variance of the energy gaps, as weighted by the absolute value of the coefficients in Eq.~\eqref{eq:aoftE}~\cite{Garcia-PintosPRX2017,de2018equilibration,Wilming17}. This exactly coincides with the expression for $\sigma_G$ in Eq.~\eqref{eq:sigmag}. This implies that the initial decay rate is the same as that of a correlation function $\corr(t)$, which we already know to be $\sigma_G$ from the discussion in Sec.~\eqref{sec:corr}, and so $\sigma_A \simeq \sigma_G$. 

However, from this coarse argument it is not clear whether a conclusion as strong as $\sigma_A \simeq \sigma_G$ can hold in full generality --- in fact, we will see that at times they are only within the same order of magnitude. Nevertheless, it suggests that this coincidence of timescales will be closer the more evenly spread the coefficients in Eq.~\eqref{eq:aoftE} are in the complex plane, and the fewer spurious correlations there are between them. We explore this numerically in Sec. \ref{sec:cete} below, and further explain it in Sec. \ref{sec:effdim}.

%%%%%%%%%%%%%%%%
%%%%%%%%%%%%%%%%%

\section{Dissipation: The Kubo thermal response function}\label{sec:kubo}

The Kubo correlation function models the dissipation of small perturbations away from thermal equilibrium~\cite{kubo1957statistical}
\begin{align}\label{eq:kubo}
\corr_{\text{Kubo}}(t) \propto \sum_{j \neq k} \frac{e^{- \beta E_j}-e^{-\beta E_k}}{E_k-E_j} \vert A_{jk}\vert^2 e^{i(E_j-E_k)t}.
\end{align}
By a similar argument as that used for $\corr(t)$ in Appendix~\ref{app:gaussian}, this function has an initial Gaussian/quadratic decay 
\begin{equation} \label{eq:kubosigma}
\corr_{\text{Kubo}}(t) \simeq \corr_{\text{Kubo}}(0) e^{-\sigma_K^2 t^2} \simeq \corr_{\text{Kubo}}(0) \left(1 -\sigma_K^2 t^2 \right),
\end{equation}
with a characteristic decay rate given by 
\begin{align}
\label{eq:kubodecay}
    \sigma_K^2 =& \sum_{j,k}\frac{e^{-\beta E_k}-e^{-\beta E_j}}{\corr_{\text{Kubo}}(0)}\vert A_{jk}\vert^2 (E_k-E_j) \\ &= \frac{1}{\corr_{\text{Kubo}(0)}} \tr{\left[A,\frac{e^{-\beta H}}{Z}\right][A,H]}.
\end{align}
The initial decay rate $\sigma_K$ also characterizes other aspects of the dynamics of the correlation function, as was the case for $\mathcal{C}(t)$. 
Theorem 5 in~\cite{alhambra2020time} shows that in some cases $\sigma_K$ also governs the equilibration timescale of $\corr_{\text{Kubo}}$, and Appendix~\ref{app:qslCkubo} shows that $\sigma_K$ also upper bounds its rate of change.

Based on the ETH, the work of Srednicki~\cite{Srednicki99} argues that $\mathcal{C}(t)$ behaves similarly to $\corr_{\text{Kubo}}(t)$. We reproduce the theoretical argument based on the ETH ans\"atz in Appendix~\ref{app:kubo}. The reason behind it is similar to that of the previous section: states that have a support uniformly spread over many energy eigenstates will behave in a more ``thermal", or ``typical" manner, so that both correlation functions evolve similarly by virtue of being largely independent of the coefficients of Eqs.~\eqref{eq:aoftE} and~\eqref{eq:kubo}. 
More specifically, if the function $f(\langle H \rangle,\omega)$ of the ETH ans\"atz [see Eq.\eqref{eq:ETH}] decays rapidly at frequencies $\omega > W$ and the initial state has variance $\lambda$, then
\begin{align}\label{eq:srekubo}
\corr(t) \simeq \corr_{\text{Kubo}}(t) +  \mathcal{O}(\beta^2 W^2) + \mathcal{O}\left(\frac{W^2}{\lambda^2}\right).
\end{align}
The function $f(\langle H \rangle,\omega)$ has been thoroughly explored in numerical simulations~\cite{Khatami13,d2016quantum,Beugeling14offdiag,Mondaini17,LeBlond_2019}, showing a relatively fast decay with frequency. This suggests that the two error terms in this equation are small, so that $\corr(t) \simeq \corr_{\text{Kubo}}(t)$ or that, at least, $\sigma_G \simeq \sigma_K$. We check this similarity for our examples in Sec. \ref{sec:cete} below. This coincidence of the dynamics of $\corr(t)$ and $\corr_{\text{Kubo}}(t)$ 
has been previously identified as a  
quantum ``fluctuation-dissipation" relation~\cite{Khatami13}.

Note that the argument leading to Eq. \eqref{eq:srekubo} crucially relies on the coefficients $c_j$ and $A_{jk}$ being uniformly distributed (see Eq.~\eqref{eq:kuboETH}). Thus, as we find, in the more ``non-generic" situations, hidden correlations between these coefficients can make the argument fail. This also agrees with the fact that only some of the examples studied in~\cite{Khatami13} show that $\corr(t) \sim \corr_{\text{Kubo}}(t)$, which correspond to the quenches in which we expect typicality. That is, when the lack of structure or correlations between the coefficients $c_j,e^{-\beta E_j}$ and $A_{jk}$, is more prominent.

%%%%%%%%%
%%%%%%%%%

\section{Comparison of timescales}
\label{sec:cete}

In the discussion above we have introduced three different rates: $\sigma_A,\sigma_G$ and $\sigma_K$, defined respectively in Eqs.~\eqref{eq:sigmaA},~\eqref{eq:sigmag} and \eqref{eq:kubodecay}, and speculated with the possibility that they may coincide in certain cases.
We now numerically compute these quantities, and explore whether this coincidence indeed happens.

\begin{figure}
 \centering
            \centering
            \includegraphics[width=0.5\textwidth]{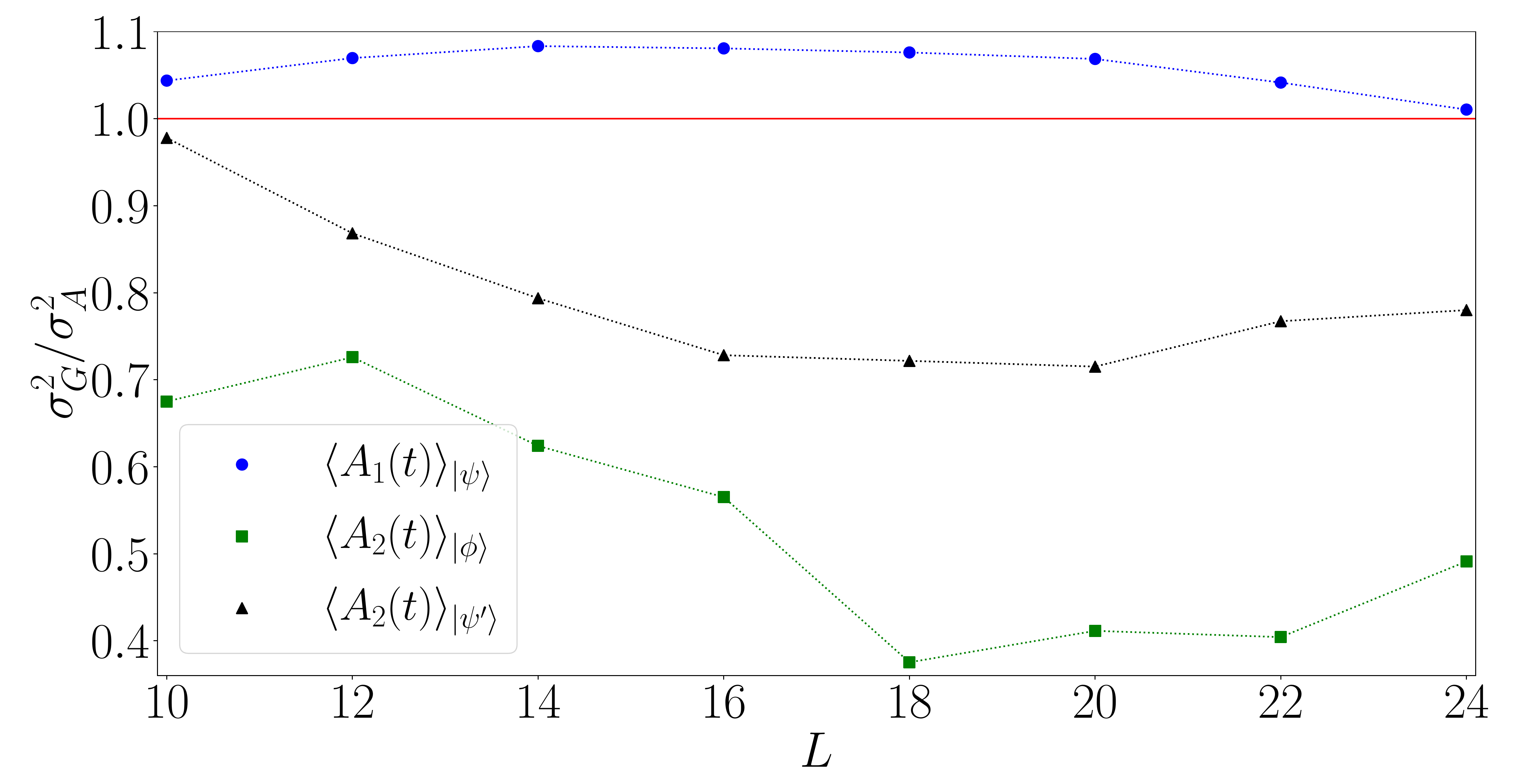}
            \caption[Network2]%
            {{\small Ratio between $\sigma_G^2$ defined in Eq.~\eqref{eq:sigmag} and $\sigma_A^2$ defined in Eq.~\eqref{eq:initialdecay}, as a function of system size. We see that in the case of $\langle A_1(t)\rangle_{\ket{\psi}}$ the ratio approaches $1$ for larger $L$. 
            }}    
           \label{fig:sigmas}
    \end{figure}

First, we compare the decay rates $\sigma_A$ and $\sigma_G$ in each of our three pairs of initial states and observables in Eqs.~\eqref{eq:obs-state}. The results are shown in Fig.~\ref{fig:sigmas}, where we see that $\sigma_A$ and $\sigma_G$ are of the same order of magnitude, and that they converge in the case of the state-observable pair $\langle A_1(t) \rangle_{\ket \psi}$ in Eq.~\eqref{eq:at} as the size of the system grows. At system size $L = 24$, 
we see the decay rates strongly coincide, with $\sigma_G^2/\sigma_A^2 \approx
1.0105$. This shows that, at least in the latter case, the two rates converge, which is consistent with our expectation from Fig.~\ref{fig:changeKappa}. Given our discussion above, we thus expect that $ \langle A_1(t) \rangle_{\ket \psi}  
$ is the most ``typical" scenario, with $ \langle A_2 (t) \rangle_{\ket \phi}$ being the least. 

The case of $\langle A_2 (t) \rangle_{\ket \phi} $ is the one for which $\sigma_A$ and $\sigma_G$ differ the most, and also the one for which $\sigma_A$ is more sensitive to changes in $\gamma_1$ (see Fig.~\ref{fig:changeKappa}). A possible reason for this is that we expect that transport processes are relevant in the relaxation of state $\ket \phi$. These are generally associated with an ``atypicality" of the dynamics, and a breakdown of random matrix theory features~\cite{Reimann_2019}, which can potentially cause correlations between the coefficients of Eq.~\eqref{eq:aoftE}.

\begin{figure}
 \centering
            \centering
            \includegraphics[width=0.5\textwidth]{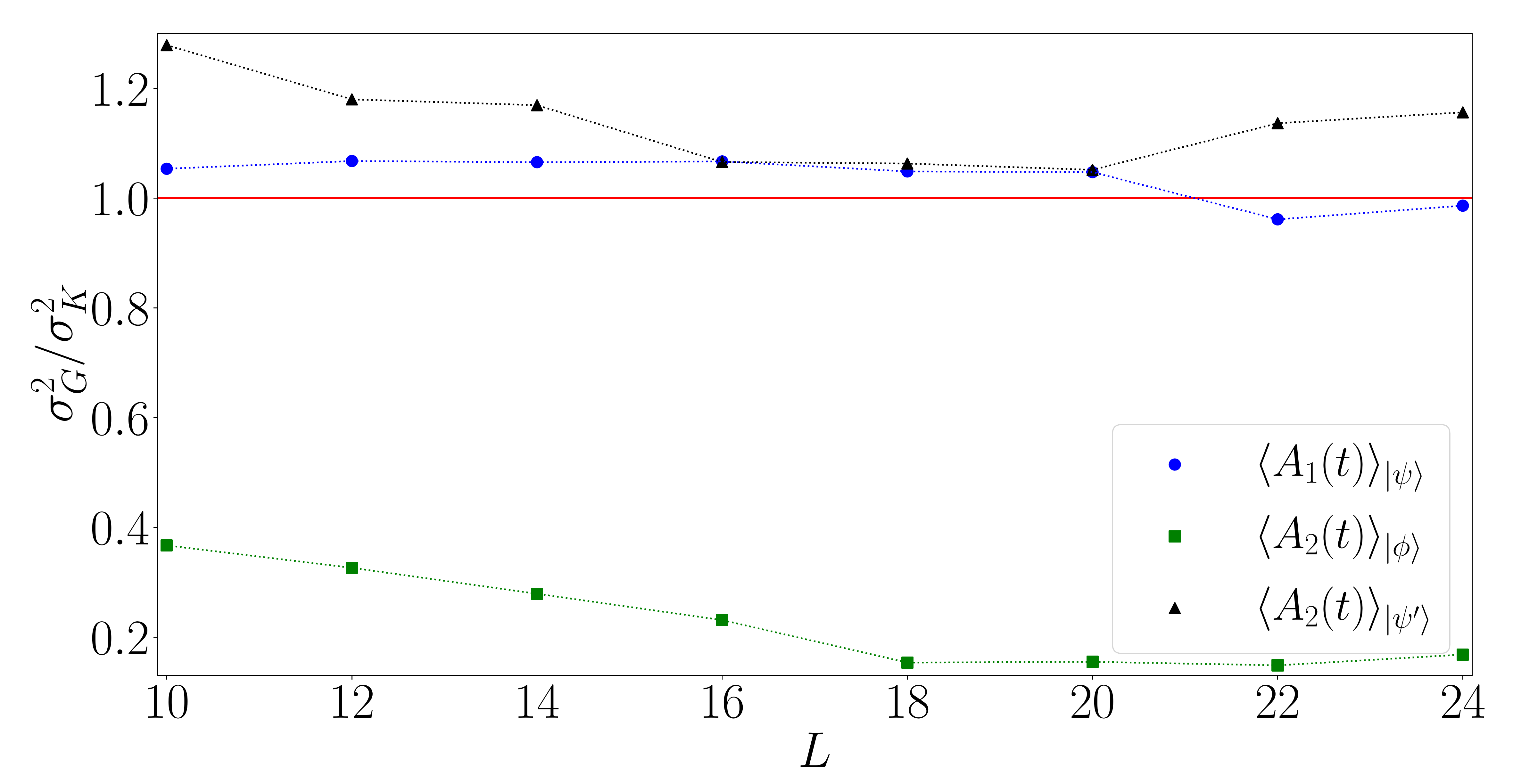}
            \caption[Network2]%
            {{\small  Ratio between $\sigma_G^2$ defined in Eq.~\eqref{eq:sigmag} and $\sigma_K^2$ defined in Eq.~\eqref{eq:kubosigma} as a function of system size. The thermal state was restricted to energy eigenstates with non-zero weight in the states defined in Eqs.~\eqref{eq:state1} and~\eqref{eq:state3}]. The value of $\beta$ was determined so that the thermal state had identical energy to the initial pure state. }}    
           \label{fig:kuboG}
    \end{figure}

Additionally, Fig.~\ref{fig:kuboG} compares the rates $\sigma_G$ and $\sigma_K$. We observe that
the two timescales are similar for both state-observable pairs $\langle A_1(t) \rangle_{\psi}$ and $\langle A_2(t) \rangle_{\phi}$, with the former being the closest. Indeed, for $\langle A_1(t) \rangle_{\ket \psi}$ and the largest system size tested $L = 24$ we have that $\sigma_G^2/\sigma_K^2\nobreak \approx\nobreak0.9864$. This shows how in some cases the aforementioned ``fluctuation-dissipation" relation can emerge in certain situations, as previously found in~\cite{Khatami13}.
On the other hand, the rates $\sigma_G$ and $\sigma_K$ differ significantly for the third case of $\langle A_2(t) \rangle_{\ket{\psi'}}$ (although still within an order of magnitude). 

Moreover, comparing Figs.~\ref{fig:sigmas} and~\ref{fig:kuboG} shows that the cases when $\sigma_K \sim \sigma_G$ coincide with those for which $\sigma_K \sim \sigma_A$. This is not surprising since, as per the discussions above, we expect that the most typical situations are the ones in which all these rates are similar. Here, the most typical case is $\langle A_1(t) \rangle_{\psi}$, with $\langle A_2(t) \rangle_{\ket{\psi'}}$ being somewhat far from it.

%%%%%%%%
%%%%%%%%

\section{Random matrix theory analysis}\label{sec:RMT}

We have seen how in the model of Eq. \eqref{eq:hamiltonian} the different timescales considered coincide for certain state-observable combinations. We now explore the conclusions of the previous sections in the paradigmatic model of ``typicality", with a Hamiltonian diagonalized by a random unitary matrix~\cite{Popescu_2006}.

This is a simpler model in which the characteristic relaxation rates can be analytically computed exactly. Let us define an arbitrary Hamiltonian in which the eigenbasis is chosen at random, as
\begin{equation}
    H_U \equiv U H U^\dagger,
\end{equation}
where $U$ is drawn from the Haar measure over the unitary group~\cite{Collins2006}. This model, which has previously appeared in the study of equilibration of closed systems \cite{Popescu_2006,Brandao12,Reimann_2015}, is motivated by the fact that quantum non-integrable systems have highly entangled eigenstates, that highly resemble random states~\cite{Deutsch91,Santos2012,Beugeling2015,d2016quantum,Vidmar2017}.

With it, we are able to give analytical expressions for the average squared decay rates $\langle \sigma^2_A \rangle_U$, $\langle \sigma^2_G \rangle_U$ and $\langle \sigma^2_K \rangle_U$, where $\langle \cdot \rangle_U$ indicates the average over the Haar measure. Given that the corresponding thermal state of this model is always the maximally mixed state, the thermal state corresponding to $\mathcal{D}(\ketbra{\Phi}{\Phi})$ is always the infinite temperature state $\beta\nobreak=\nobreak 0$.

In Appendix \ref{app:random}, we show that 
\begin{align}\label{eq:randomrates}
    \langle \sigma^2_G \rangle_U & \simeq \langle \sigma^2_A \rangle_U + \mathcal{O}\left(\frac{1}{d} \right)= \langle \sigma^2_K \rangle_U + \mathcal{O}\left(\frac{1}{d} \right) \\& \nonumber = 2 \left(\langle H^2 \rangle_{\text{MC}}- \langle H \rangle_{\text{MC}}^2\right)+ \mathcal{O}\left(\frac{1}{d} \right),
\end{align}
where $d = 2^L$ is the dimension of the total Hilbert space and $\langle B \rangle_{\text{MC}} \nobreak=\nobreak \tr{B}/d$.  That is, the rates coincide with the energy variance at infinite temperature for $L \gg 1$. The intuitive reason behind this is that the randomness in the eigenstates washes out any correlations between $\{A_{jk}\}$, $\{c_j,c_k\}$ and $E_j-E_k$, such that the relevant timescale does not depend on the observable nor the initial state but only on the spectrum. The $``\simeq"$ in Eq. \eqref{eq:randomrates} relies on the accuracy of the approximation $\langle f/g \rangle_U \approx \langle f \rangle_U / \langle g \rangle_U$, sometimes referred to as the ``annealed approximation"~\cite{meir1995stochastic,PhysRevE.72.061905,cotler2017black,cotler2017chaos,ChenuQuantum2019,PhysRevLett.122.014103,PhysRevX.10.031026}, which we justify analytically in Appendix~\ref{app:annealed}.

To derive Eq.~\eqref{eq:randomrates}, we need to compute the average of certain correlation functions over the unitary group. These involve the fourth moment of the Haar measure, which results in cumbersome analytical expressions of hundreds of coefficients coming from the Weingarten calculus~\cite{Collins2006}. We deal with these analytically with the recently introduced RTNI package~\cite{RTNI2019}. In contrast, the calculations with the same model of e.g.,~\cite{reimann2016typical} only involve second moments, which can be done by hand.

The result in Eq. \eqref{eq:randomrates} supports the fact that, in typical  instances of the dynamics where RMT is accurate, the timescales studied here coincide with a ``universal'' value. It is also consistent with previous studies of RMT models~\cite{reimann2016typical,nation2018quantum}, where such coincidence of timescales for different dynamical processes is already hinted at.

%%%%%%%%%%
%%%%%%%%%

\section{Typicality of the dynamics and the effective dimension} \label{sec:effdim}

Here, we propose to quantify the typicality of the situations studied with two figures of merit, and connect them to the dynamics of the observables and correlation functions.
 
The dephasing argument in Sec.~\ref{sec:corr} suggests that the more ``evenly spread" in frequency the coefficients $c_jc_k^* A_{jk}$ are, the more we expect the decay rates to coincide, $\sigma_A \sim \sigma_G$. On top of that, the result of the Sec.~\ref{sec:RMT} shows that, when these are fully random, the rates coincide for large Hilbert space dimension $d$.  Motivated by these facts, we aim to understand our numerical findings from Sec.~\ref{sec:cete} in terms of two different measures of typicality of the initial conditions.

The first one contains information about the state and the Hamiltonian, and is the so-called \emph{effective dimension}, also commonly referred to as the inverse participation ratio. The effective dimension $D_{\Phi} $ of a pure state $\ket{\Phi}$,  
defined by
\begin{equation}
D^{-1}_{\Phi} \equiv \tr{\mathcal{D}(\ketbra{\Phi}{\Phi})^2}=\left(\sum_j \vert c_j\vert^4 \right),
\end{equation}
 controls the long-time equilibration of isolated quantum systems~\cite{Popescu05,Reimann08,Lin09}. Here, $\overline{\ket{\Phi(t)}\!\bra{\Phi(t)}}=\mathcal{D}(\ketbra{\Phi}{\Phi})$ is the diagonal ensemble: the initial state dephased in the energy eigenbasis. 

The effective dimension quantifies
the number of eigenstates that participate in the process. A ``typical" state thus has a large effective dimension. It is known to grow exponentially in system size under very general conditions on the eigenstates~\cite{wilming2018entanglement,rolandi2020extensive} and to be close to the maximal value $d$ if the initial state is chosen at random~\cite{huang2020instability,haferkamp2021emergent}. Importantly, it bounds the size of late time fluctuations around equilibrium~\cite{Reimann08,short2011equilibration}. 

A shortcoming of this measure is that it does not depend on the observable studied.
 To account for it, we consider a second measure of typicality $D_{\Phi,A}$ that incorporates information of the off-diagonal matrix elements of the observable, which play a role in dynamics. We refer to $D_{\Phi,A}$ as a \emph{state-observable effective dimension}, and define it by \footnote{Notice that this quantity is very close to $\tr{A\mathcal{D}(\ketbra{\Phi}{\Phi})A\mathcal{D}(\ketbra{\Phi}{\Phi})}$ since they differ only by the $j=k$ terms.}
\begin{equation} \label{eq:Np}
     D^{-1}_{\Phi,A} \equiv  \sum_{j\neq k} |c_j|^2 |c^*_k|^2 |A_{jk}|^2.
\end{equation}

It holds that $D^{-1}_{\Phi,A} \leq \vert \vert A \vert \vert^2 D^{-1}_{\Phi}$ \cite{Reimann08,short2011equilibration}, where $\vert \vert A \vert \vert$ is the largest singular value of $A$. 
Comparing  $D^{-1}_{\Phi,A}$ and $D^{-1}_{\Phi}$, we can roughly think of $D^{-1}_{\Phi,A}$ as an inverse effective dimension that accounts for the observables off-diagonal elements. Notice that this expression only contains information about the off-diagonal terms which generate the dynamics, and not the diagonal ones.

The state-observable effective dimension is related to the fluctuations around equilibrium, as
\begin{equation}
D^{-1}_{\Phi,A} = \lim_{T\to \infty} \int_0^\infty \frac{\text{d}t}{T} |\langle A (t) \rangle - \tr{\mathcal{D}\left(|\Phi\rangle \langle \Phi | \right) A} |^2 ,
\end{equation}
where we have assumed non-degeneracy in the energy gaps.

The significance of these two measures for the dynamics at earlier times can be understood as follows: the larger the effective dimensions, the more off-diagonal terms $A_{jk}$ can participate in the dynamics. This implies that the particular details and structure of individual off-diagonal matrix elements contribute less at an earlier time, making treatments akin to random-matrix theory, such as in~\cite{reimann2016typical,reimann2019transportless,richter2018impact,Nation_2019}, more accurate. That is, we generally expect that the larger the effective dimensions $D_{\Phi}$ and $D_{\Phi,A}$, the more typical the dynamics is.

When that is the case, the initial state and the energy eigenbasis are closer to being ``mutually unbiased"~\cite{Anza_2018}, which effectively means that the energy eigenbasis and a basis of low-entangled states including $\ket{\Psi}$ can be related by a random matrix. In \cite{reimann2016typical}, this was shown to imply that
\begin{align}\label{eq:formfactor}
\langle A(t)\rangle \propto \sum_{j,k} e^{-it(E_j-E_k)},
\end{align}
where the sum includes the set of energy gaps that participate in the dynamics. That is, $\langle A(t)\rangle$ resembles the \emph{spectral form factor} of the Hamiltonian, and depends weakly on the details of the coefficients $A_{jk}$ and $c_j$. This is consistent with the random matrix theory result of Section~\ref{sec:RMT}, which show that in a random-matrix model the timescales are set only by the Hamiltonian. 
 
Another way to understand this is as follows: if we compute the coefficients $c_j,A_{jk}$ with a random matrix formalism, they will be all of similar weight, and close to maximally evenly distributed in the complex plane, leading on expectation to Eq.~\eqref{eq:formfactor}.

In contrast, in a local model with energy conservation, such as that of Eq. \eqref{eq:hamiltonian}, the coefficients $A_{jk}$ and $c_j$ have non-trivial structure, and set the range of participating energy gaps to be those around the average energy (those for which $c_j c_k^* A_{jk}$ is not too small). Due to this, the decay rates $\sigma_A$ can be significantly different from that of an actual form factor or the Loschmidt echo, which would correspond to the energy variance~\cite{Campos_Venuti_2010,Alhambra_2020} as we found in Eq.~\eqref{eq:randomrates}. 

Exactly the same argument applies to $\corr(t)$ and $\corr_\text{Kubo}(t)$, and one can thus infer that they will behave similarly to $\langle A(t) \rangle$ on the basis of them also evolving like an spectral form factor $\sum_{j,k} e^{-it(E_j-E_k)}$ of the participating energy gaps. This idea is consistent with~\cite{cotler2019spectral}, which proposes that the coincidence of correlation functions and form factors is a key indicator of the validity of random matrix theory ans\"{a}tze.

\begin{figure}
 \centering
            \centering
            \includegraphics[width=0.5\textwidth]{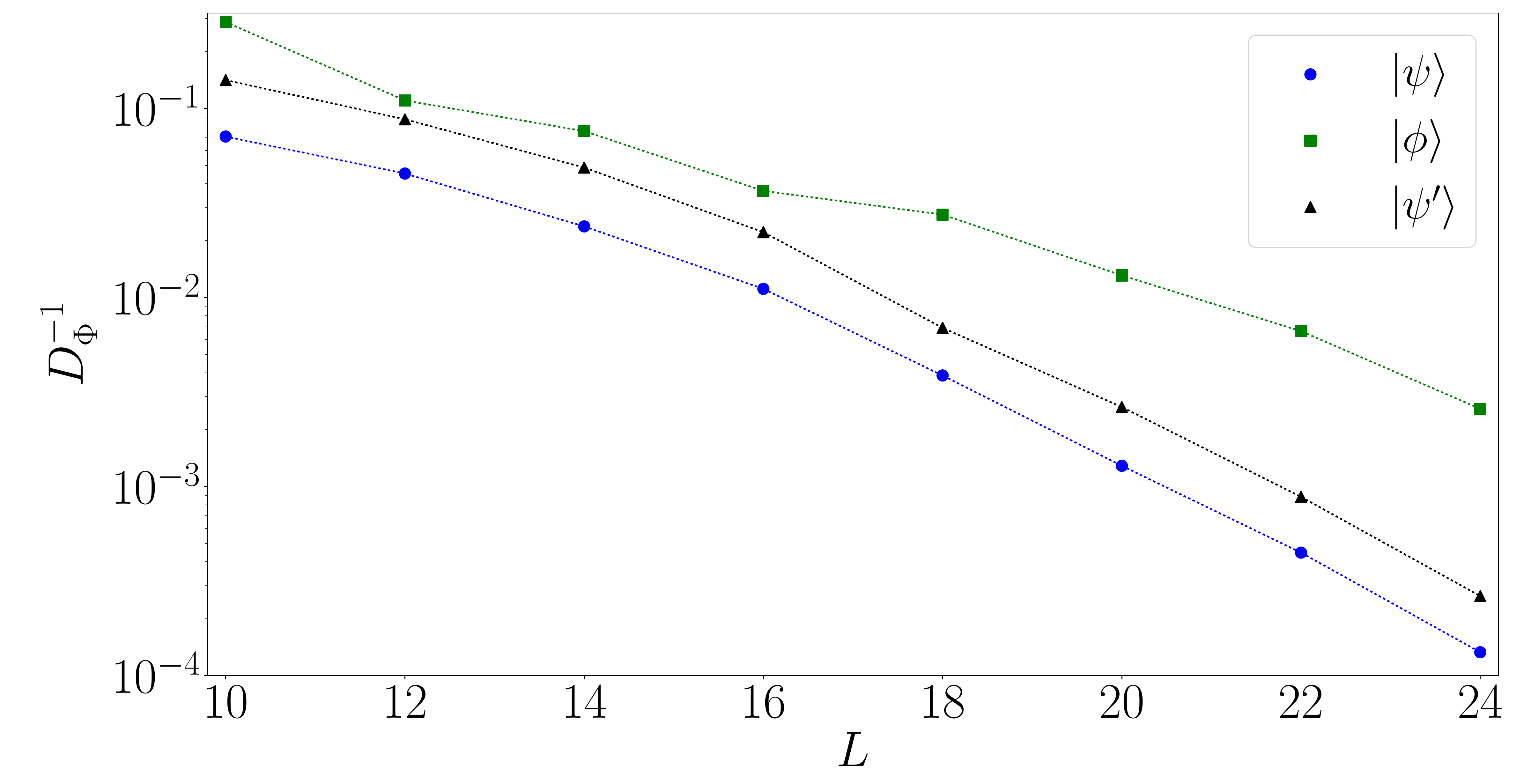}
            \caption[Network2]%
            {{\small Inverse effective dimension $D^{-1}_{\Phi}$ as a function of system size, up lo $L=24$. We see a clear exponential decay, with different states decaying at diverging rates.
             }}    
           \label{fig:effdim}
    \end{figure}

\begin{figure}
 \centering
            \centering
            \includegraphics[width=0.5\textwidth]{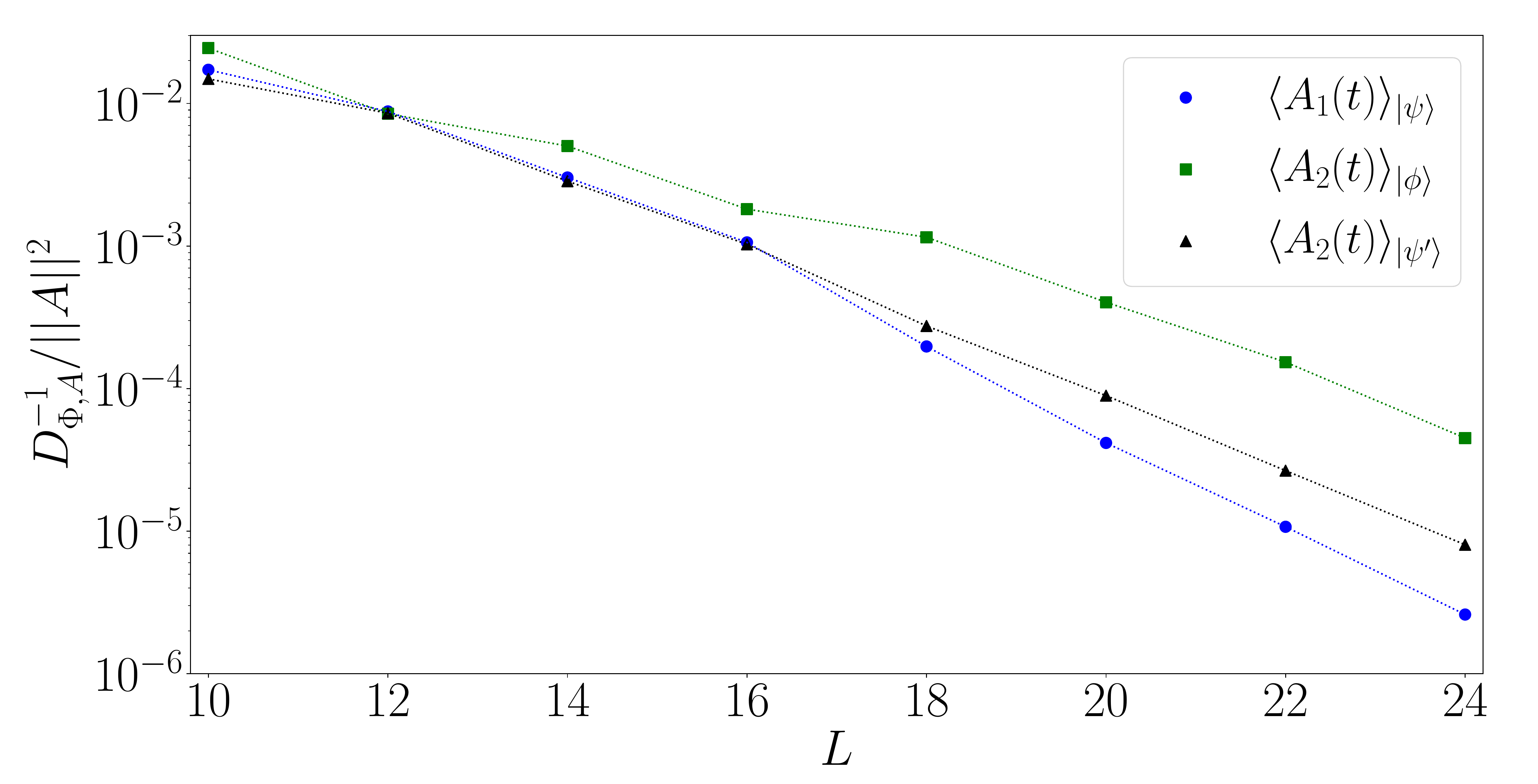}
            \caption[Network2]%
            {{\small 
            $D_{\Phi,A}^{-1}$ as a function of system size, up to $L = 24$. We see a clear exponential decay setting in at large system sizes at different rates.
            }}    
           \label{fig:Np}
    \end{figure}

Given this discussion, we expect that the larger the effective dimensions $D_{\Phi}$ and $D_{\Phi,A}$, the closer the decay rates $\sigma_A$, $\sigma_G$ and $\sigma_\textnormal{Kubo}$ shown in Fig.~\ref{fig:sigmas} of Sec.~\ref{sec:cete} become. We confirm this by calculating the effective dimensions as a function of system size for the three different initial states. The results are shown in Figs.~\ref{fig:effdim} and~\ref{fig:Np}, where we see an exponential decay in all cases. We also observe a noticeable difference between the different states, with $D_{\psi}> D_{\psi'} >D_{\phi}$, and with $D_{\psi,A_1}> D_{\psi',A_2} >D_{\phi,A_3}$ for sufficiently large systems with $L \geq 16$. This agrees with our expectations that the effective dimensions can serve to witness the situations in which the decay rates coincide. This is also consistent with the results of Sec.~\ref{sec:cete} and with Figs.~\ref{fig:sigmas} and~\ref{fig:kuboG}. We find that $\langle A_1(t) \rangle$ is the most typical scenario, and the one in which the timescales $\sigma_A,\sigma_G,\sigma_K$ most resemble each other. In contrast, $\langle A_2 (t) \rangle_{\ket \phi}$ is a more atypical scenario in which the decay rates differ the most.

These results suggest that the effective dimensions considered can serve as
\begin{itemize}
    \item Figures of merit of the typicality of the dynamics, and of the validity of RMT frameworks.
    \item Indicators of the coincidence of the relaxation rate of the quench with correlation functions.
\end{itemize}
 An interesting open problem is to make this connection more precise and quantitative, perhaps in the form of a bound similar to those for late-time fluctuations in \cite{short2011equilibration,Reimann08,Reimann10}. 


%%%%%
%%%%%

%%%%%%%%%%%%%%%%%%%%%
%%%%%%%%%%%%%%%%%%%%%

 \section{Conclusions}
 \label{sec:conclusion}
 
 We have analyzed the early time relaxation rate of observables with different initial conditions for a non-integrable quantum system. As a general conclusion, we observe a close link between the coincidence of different dynamical quantities and their initial relaxation rates with the typicality of the initial conditions. We quantify this typicality with two notions of ``effective dimensions'' that, in broad terms, quantify the number of frequencies involved in the evolution of an observable. This allows us to better understand how and when the complex relaxation behaviour of these quantum systems can be understood in simpler terms, either by linking it to correlation functions, or by having random matrix theory/typicality treatments to accurate describe dynamics. 
 
 We see that, in typical situations, the rate $\sigma_A$ at which an observable initially decays is related to the rate $\sigma_K$, which dictates the approach to thermal equilibrium from perturbations, and to the rate $\sigma_G$, which dictates temporal fluctuations. This suggests a connection between (short-time) equilibration, fluctuation, and dissipation processes of an observable for typical states -- a sort of \emph{equilibration-fluctuation-dissipation} relation, formalized by $\sigma_A \simeq \sigma_K \simeq \sigma_G$, that goes beyond some previously studied fluctuation-dissipation relations~\cite{Khatami13}.

 A full theoretical characterization of  relaxation dynamics and their timescales remains a largely open problem \cite{Wilming_2018,heveling2020compelling,Knipschild2020}.  Our findings suggest that, while it may be possible to describe the dynamics of very generic situations, the existence of ``atypical" situations make general rigorous results challenging~\cite{Goldstein13,Malabarba14,kim2015slowest}. 
 This impacts the 
 regimes in which previous upper bounds on equilibration timescales correctly capture the dynamics of isolated systems~\cite{Garcia-PintosPRX2017,de2018equilibration}. An important milestone could be to find more systematic ways of knowing when a particular dynamics can be considered typical, via the effective dimensions proposed here, or some related figure of merit.

 The present results confirm the intuition provided by a large number of previous theoretical works that make similar connections through either typicality arguments~\cite{Popescu05,Gemmer2009,reimann2016typical} or random matrix theory ans\"atze~\cite{Deutsch91,Brandao12,masanes2013complexity,nation2018off,Nation_2019}. They are also consistent with the picture that links typicality and the validity of different random matrix formalisms to the absence of macroscopic transport phenomena~\cite{Reimann_2019}.
 
 \ 
 
 \acknowledgements
 
 AMA acknowledges support from the Alexander von Humboldt foundation. J.R acknowledges support from the Natural Sciences and Engineering Research Council of Canada (NSERC). J.R would also like to thank Steven Silber and Erik S\o rensen for helpful discussions about exact diagonalization methods. 
LPGP acknowledges the DoE ASCR Accelerated Research in Quantum Computing program (award No. DE-SC0020312), DoE QSA, NSF QLCI, NSF PFCQC program, U.S. Department of Energy Award No. DE-SC0019449, DoE ASCR Quantum Testbed Pathfinder program (award No. DE-SC0019040), AFOSR, ARO MURI, AFOSR MURI, and DARPA SAVaNT ADVENT.

 \bibliography{references}
 
\clearpage 
 \widetext
\appendix
\part*{ 
 \begin{center}
 \normalsize{
 APPENDICES 
 } 
 \end{center}
 }

\section{Initial Gaussian decay} \label{app:gaussian}

Here we explain why we expect $\corr(t)$ to decay as a Gaussian for early times, which also implies that it is well approximated by a quadratic function. Similar results should also apply for $\mathcal{C}_{\text{Kubo}}$, as well as for the central quantity $\langle A (t)\rangle$. For the latter, however, the complex coefficients in the expansion Eq. \eqref{eq:aoftE} make that analysis far from straightforward.

We follow a method for estimating expectation values based on ETH that can be found in several references \cite{Khatami13,d2016quantum,murthy2019bounds,foini2019eigenstate}, and that allows us to transform sums over energy gaps into integrals over frequencies. First, notice the dependence of Eq. \eqref{eq:coft} on the matrix elements $A_{jk}$. We can then make use of the ETH ansatz 
 \begin{align}\label{eq:ETH}
&\bra{E_j} A \ket{E_k}\equiv A_{jk}=A(E)\delta_{jk}+e^{-S(E)/2}f(E,E_j-E_k)R_{jk}. 
\end{align}
 Here, $E_j,E_k$ are energies belonging to the same microcanonical ensemble with energy $E$. $S(E)$ is the microcanonical entropy of that ensemble, $f(E,\omega)$ is a function that decays monotonically  with $\omega$ and $R_{jk}$ are the coefficients of a random matrix.

The randomness in Eq. \eqref{eq:ETH}, and the small level spacing in the thermodynamic limit, suggests that we can replace the sums $\sum_{k,j} \vert c_j \vert^2 \vert c_k \vert^2$ in \eqref{eq:coft} with integrals over the energy sum and differences $ \int\int \text{d} E \text{d} \omega e^{\beta_E S(E+\omega)} p(E-\omega/2) p(E+\omega/2)$ \cite{d2016quantum}. Here $e^{\beta_E S(E)}$ is the density of states at energy $E$, and $\beta_E$ is the inverse temperature corresponding to average energy $E$. The integral is as follows
\begin{align} \label{eq:integralC}
\mathcal{C}(t) - \mathcal{C}(\infty) \propto & \int \int \text{d} E \text{d} \omega  e^{\beta_E (S(E+\omega)-S(E))}   p(E-\omega/2) p(E+\omega/2) \vert f(E,\omega) \vert^2 e^{i \omega t},
\end{align}
where $p(E)$ is the probability of being in energy $E$, the continuum limit of the coefficients $\vert c_j\vert^2$, and $\mathcal{C}(\infty) \nobreak = \nobreak \overline \corr(t)$.

For physically relevant initial states, such as product or shortly-correlated states on lattices, this energy distribution is always close to a Gaussian (see \cite{operatorid1,brandao2015equivalence} for rigorous statements)
\begin{align}\label{eq:gaussian}
p(E) \simeq \frac{1}{\sqrt{2 \pi \lambda}}e^{-\frac{(E-\langle H \rangle)^2}{2 \lambda^2}},
\end{align}
where $\langle H \rangle = \bra{\Phi} H \ket{\Phi}$ and $\lambda^2=\bra{\Phi} H^2 \ket{\Phi}-(\bra{\Phi} H \ket{\Phi})^2$. Typically, $\lambda^2 \sim N$, so that the energy fluctuations are subextensive, and thus the energy density is essentially free of uncertainty. We can then write 
\begin{equation}
p(E+\omega/2)p(E-\omega/2) \propto e^{-\frac{(E-\langle H \rangle)^2}{\lambda^2}} e^{-\frac{\omega^2}{2 \lambda^2}}.
\end{equation}

This means that the energy is highly peaked around the average value, with small fluctuations around it. This fixes the ``effective" temperature $\beta_E= \beta_{\langle H \rangle}\equiv \beta$. At the same time,  $S(E+\omega)$ can only change significantly if $\omega$ changes by an extensive amount. As such, we can approximate $S(E+\omega)-S(E)$ to leading order in $\omega$, obtaining
\begin{equation}\label{eq:dofe}
 e^{\beta_E (S(E+\omega)-S(E))} \simeq e^{\beta_E \frac{\omega}{2}} = e^{\beta \frac{\omega}{2}}.
\end{equation}

Notice that the only dependence left on $E$ is on the function $f(E,\omega)$, which again changes very slowly with $E$ (it should be effectively constant within the same energy density). Putting everything together, we can write
\begin{align}
\label{eq:finalct}
\corr(t) - \corr(\infty) &\propto \int \text{d} \omega e^{\beta \frac{\omega}{2}}  \vert f(\langle H \rangle, \omega) \vert^2 e^{i t \omega} e^{-\frac{\omega^2}{2 \lambda^2}} \nonumber \\
& \propto \int \text{d} \omega  \vert f(\langle H \rangle, \omega) \vert^2 e^{i t \omega} e^{-(\frac{\omega}{\sqrt{2}\lambda}-\frac{\beta \lambda}{\sqrt{8}})^2}.
\end{align}

An extra constraint on the observable is imposed by its locality. This implies that $\tr{\rho A^2}$ must be finite and $\mathcal{O}\left(1\right)$ for any $\rho$, and it was argued in \cite{murthy2019bounds} [Eq.~(12)] that this implies that at large enough $\omega$, $f$ decays at least as fast as
\begin{equation}\label{eq:decayf}
f(\langle H \rangle ,\omega) \sim e^{-\frac{\beta \vert \omega \vert}{4}}.
\end{equation}
This exponential decay at large frequencies has been numerically verified in at least~\cite{Khatami13,d2016quantum,Beugeling14offdiag,Mondaini17,LeBlond_2019} (see \cite{Arad16,de2018equilibration} for mathematically rigorous but weaker statements).
It further justifies the approximation in Eq. \eqref{eq:dofe}, since the integrand will be very suppressed at large $\omega$.
Essentially, this means that if we are interested in short times $t$, for which the high frequencies matter more, we can assume that $f(E,\omega)$ decays like a simple exponential in $\vert \omega \vert$. We do not have an estimate for the cut-off frequency or time at which this argument starts to fail, but it is likely some timescale related to the transport processes in the system. On the other hand, $f(\langle H \rangle ,\omega)$ is roughly constant for some range of small frequencies $\omega$. This can be thought of as a consequence of the validity of random matrix theory-like phenomena at late times, after all transport phenomena have dissipated~\cite{Dymarsky2018bound,Richter_2020,brenes2021outoftimeorder}.

We see then that the integrand \emph{at short times} is just a product of Gaussians and of decaying exponentials, with the Fourier factor $e^{i \omega t}$. We conclude that 
\begin{equation}\label{eq:gaussianI}
\int \text{d} \omega e^{- \frac{(\omega-\omega_0)^2}{2 \sigma^2}} e^{i t \omega} \propto e^{- \sigma^2 (t-t_0)^2 },
\end{equation}
where $\omega_0$ and $\sigma$ depend on the details of the function $f$, and on the constants $\beta$ and $\lambda$. 

At the same time, we can see that if we Taylor-expand $\corr(t)$ and write it as in Eq. \eqref{eq:coft}, we have
\begin{align}
\corr(t) &= 1 - \frac{t^2}{2} \frac{\sum_{j, k} \vert c_j \vert^2 \vert c_k\vert ^2 \vert A_{jk} \vert^2 (E_k-E_j)^2}{\sum_{j, k} \vert c_j \vert^2 \vert c_k\vert ^2 \vert A_{jk} \vert^2 }+\mathcal{O}\left(t^3 \right) \nonumber \\ & = 1-\frac{\sigma_G^2 t^2}{2}+ \mathcal{O} \left(t^3 \right ),
\end{align}
where we have defined
\begin{equation}\label{eq:sigmag2}
    \sigma_G^2 \coloneqq \frac{\sum_{j, k} \vert c_j \vert^2 \vert c_k\vert ^2 \vert A_{jk} \vert^2 (E_k-E_j)^2}{\sum_{j, k} \vert c_j \vert^2 \vert c_k\vert ^2 \vert A_{jk} \vert^2 }.
\end{equation}
Given the form of \eqref{eq:gaussianI}, this shows that the initial decay constant is $\sigma=\sigma_G$ and $t_0=0$, which is confirmed by our numerical examples.

A similar argument  can be also done for the Gaussianity of the early-time decay of $\langle A(t) \rangle$ (e.g. see \cite{Wilming17}), with the extra potential difficulty of the \emph{relative phases} of the complex $c_j,A_{jk}$. Our numerical calculations support the conclusion that both $\corr(t)$ and $\langle A(t) \rangle$ decay as Gaussians at early times.

\section{Upper bound on the rate of change of $\corr(t)$}\label{app:qslC}

From Eq.~\eqref{eq:coft} in the main text we have that
\begin{align}
\left| \frac{d \corr(t)}{dt} \right| = \frac{1}{\sum_{jk} |c_j|^2 |c_k|^2 |A_{jk}|^2} \left| \sum_{j\neq k} \vert c_j \vert^2 \vert c_k\vert ^2 \vert A_{jk} \vert^2 e^{i t (E_j-E_k)} (E_j-E_k) \right|.
\end{align}
Using the Cauchy-Schwarz inequality and Eq.~\eqref{eq:sigmag} in the main text gives that
\begin{align}
\left| \sum_{j\neq k} \vert c_j \vert^2 \vert c_k\vert ^2 \vert A_{jk} \vert^2 e^{i t (E_j-E_k)} (E_j-E_k) \right|^2 &\leq  \sum_{j\neq k} \vert c_j \vert^2 \vert c_k\vert ^2 \vert A_{jk} \vert^2            \sum_{j\neq k} \vert c_j \vert^2 \vert c_k\vert ^2 \vert A_{jk} \vert^2   (E_j-E_k)^2 \nonumber \\
&\leq  \left( \sum_{j k} \vert c_j \vert^2 \vert c_k\vert ^2 \vert A_{jk} \vert^2    \right)^2\sigma_G^2.
\end{align}
Thus, 
\begin{align}\label{eq:boundrate}
    \left| \frac{d \corr(t)}{dt} \right| \leq \sigma_G.
\end{align}
This bounds the rate of change of $\corr(t)$ by its short time decay rate $\sigma_G$, and this holds at all times $t$.

\section{Upper bound on the equilibration timescale of $\corr(t)$}\label{app:equilibrationC} 
 
Defining $\corr(\infty) \coloneqq \overline{\corr(t)} = \mathcal{D}(\ket{\Phi}\bra{\Phi})$, we have 
\begin{align}\label{eq:cete}
\corr(t)-\corr(\infty) &= \frac{\sum_{j\neq k} \vert c_j \vert^2 \vert c_k\vert ^2 \vert A_{jk} \vert^2 e^{i t (E_j-E_k)} }{\sum_{j, k} \vert c_j \vert^2 \vert c_k\vert ^2 \vert A_{jk} \vert^2 } \\&= \frac{\sum_{j\neq k} \vert c_j \vert^2 \vert c_k\vert ^2 \vert A_{jk} \vert^2 e^{i t (E_j-E_k)} }{\tr{\mathcal{D}(\ket{\Phi}\bra{\Phi}) A \mathcal{D}(\ket{\Phi}\bra{\Phi}) A}}.
\end{align}
Let us denote the normalized distribution $q_\alpha \coloneqq \tfrac{1}{K}\vert c_j \vert^2 \vert c_k\vert ^2 \vert A_{jk} \vert^2$, where $\alpha = (j,k)$ denotes pairs of energy levels and $K\coloneqq\tr{\mathcal{D}(\ket{\Phi}\bra{\Phi}) A^\dagger \mathcal{D}(\ket{\Phi}\bra{\Phi}) A}$. Then, we can write
\begin{align}
\langle |\corr(t)-\corr(\infty)|^2 \rangle_T &= \sum_{\alpha,\beta} q_\alpha q_\beta \left\langle e^{-it (G_\alpha-G_\beta)} \right\rangle_T,
\end{align}
where $\langle f(t) \rangle_T \coloneqq \frac{1}{T} \int_0^T f(t) dt $ denotes time average.

Lemma 2 of~\cite{alhambra2020time} and Proposition 5 of~\cite{Garcia-PintosPRX2017}  imply that
\begin{align}
\langle |\corr(t)-\corr(\infty)|^2 \rangle_T & \leq 3\pi \left( \frac{a(\epsilon) }{\sigma_G T} + \delta(\epsilon)\right). 
\end{align}
Here, $a(\epsilon)$ and $\delta(\epsilon)$ are functions of energy gaps that depend on the form of the distribution $q_\alpha$, and therefore depend on the observable, initial state, and Hamiltonian of the system. One can argue that, typically, one can find $\epsilon$ such that $a(\epsilon) \sim 1$ and $\delta(\epsilon) \ll 1$ for generic many-body systems (see~\cite{Garcia-PintosPRX2017,alhambra2020time} for more details, and~\cite{heveling2020compelling} for a discussion of cases when this condition may not hold). 
 
 The variance of the energy gaps $G_\alpha$ with respect to the distribution $q_\alpha$ is given by
\begin{align}
\label{eq:sigmaSrednicki}
\sigma^2_{G} &=\frac{1}{K} \sum_{jk} \vert c_j \vert ^2 \vert c_k \vert ^2 \vert A_{jk} \vert ^2 (E_j-E_k)^2
\\&= \frac{\tr{\mathcal{D}(\ket{\Phi}\bra{\Phi}) [A,H] \mathcal{D}(\ket{\Phi}\bra{\Phi}) [H,A]}}{\tr{\mathcal{D}(\ket{\Phi}\bra{\Phi}) A \mathcal{D}(\ket{\Phi}\bra{\Phi}) A}},
\end{align}
and when $a(\epsilon) \sim 1$ it dominates the approach to equilibrium of $\corr(t)$. Note that this matches the short-time decay rate (Eq. \eqref{eq:sigmag})  as well as the fastest rate of change of the correlation function (Eq. \eqref{eq:boundrate}).

\section{The Kubo function and $\mathcal{C}(t)$}\label{app:kubo}
 
Making use of the ETH ansatz as in Eq. \eqref{eq:integralC}, we obtain that, for the Kubo function in Eq.~\eqref{eq:kubo},
\begin{align}\label{eq:kuboETH}
\corr_{\text{Kubo}}(t) \propto &\int \text{d}E \int \text{d} \omega e^{\beta_E S(E+\omega)}e^{-\beta_E S(E)}  \vert f(E, \omega)\vert^2 e^{i \omega t} \frac{\sinh (\frac{\beta \omega}{2})}{\omega} e^{-\beta E}.
\end{align}
The integrand is thus proportional to the Gibbs distribution $e^{-\beta E}$. In most situations, this is very peaked around the average energy $\langle H \rangle$~\cite{anshu2016concentration}, in the same way as above for the initial pure states are (since both are states with short-range correlations). This means that the typical energy fluctuations are subextensive, and the system has a well defined energy density. As such, we can treat it in the same way as the diagonal distribution in Eq.~\eqref{eq:gaussian}: effectively a Dirac $\delta$ function centered at the average energy $\langle H \rangle$. Moreover, as in Eq.~\eqref{eq:dofe}, we can approximate
$e^{\beta_E (S(E+\omega)-S(E))} \simeq e^{\beta_E \frac{\omega}{2}}$ and write
\begin{align}
\corr_{\text{Kubo}}(t) \propto \int \text{d} \omega e^{\beta \frac{\omega}{2}} \vert f(\langle H \rangle, \omega)\vert^2 e^{i \omega t} \frac{\sinh (\frac{\beta \omega}{2})}{\omega}.\label{eq:finalkubo}
\end{align}
Notice that the difference between Eq. \eqref{eq:finalkubo} and Eq. \eqref{eq:finalct} is only on the last factor, and that for small $\omega$,
\begin{align}
\frac{\sinh (\frac{\beta \omega}{2})}{\omega} \simeq 1 + \mathcal{O}(\beta^2 \omega^2)
\\
e^{-\frac{\omega^2}{2 \lambda^2}} \simeq 1+ \mathcal{O}\left(\frac{\omega^2}{\lambda^2}\right).
\end{align} 
Thus, if the function $f$ decays quickly on an energy scale of $\omega \sim W$, we expect that 
\begin{equation}
\corr(t) \simeq \corr_{\text{Kubo}}(t) +  \mathcal{O}(\beta^2 W^2) + \mathcal{O}\left(\frac{W^2}{\lambda^2}\right).
\end{equation}
This is the conclusion of~\cite{Srednicki99}. That this is the case, and that these two functions coincide, has been verified in at least~\cite{Khatami13,Mondaini17,LeBlond_2019}. There, it is shown that $f(E,\omega)$ is constant for some small interval around $\omega=0$, and then quickly decays in an exponential fashion. We provide further evidence within our setting in Sec.~\ref{sec:cete}.

\section{Upper bound on the rate of change of $\corr_{\text{Kubo}}(t)$}\label{app:qslCkubo}

From Eq.~\eqref{eq:kubo} in the main text, we have that the rate of change of the Kubo correlation function satisfies
\begin{align} 
\left| \frac{d \, \corr_{\text{Kubo}}(t)}{dt} \right| = \frac{1}{\corr_{\text{Kubo}}(0)} \left| \sum_{j \neq k} \frac{e^{- \beta E_j}-e^{-\beta E_k}}{E_k-E_j} \vert A_{jk}\vert^2  e^{i(E_j-E_k)t} (E_j-E_k) \right|.
\end{align}
The Cauchy-Schwarz inequality implies that
\begin{align} 
\left| \sum_{j \neq k} \frac{e^{- \beta E_j}-e^{-\beta E_k}}{E_k-E_j} \vert A_{jk}\vert^2  e^{i(E_j-E_k)t} (E_j-E_k)  \right|^2 &\leq
\left| \sum_{j \neq k} \frac{e^{- \beta E_j}-e^{-\beta E_k}}{E_k-E_j} \vert A_{jk}\vert^2   \right| 
\, 
\left| \sum_{j \neq k} \frac{e^{- \beta E_j}-e^{-\beta E_k}}{E_k-E_j} \vert A_{jk}\vert^2  (E_j-E_k)^2  \right| \nonumber \\
&= \corr_{\text{Kubo}}(0) \, 
\left| \sum_{j \neq k} \Big( e^{- \beta E_j}-e^{-\beta E_k} \Big) \vert A_{jk}\vert^2  (E_j-E_k) \right| \nonumber \\
&= \corr_{\text{Kubo}}^2(0) \, \sigma_K^2,
\end{align}
where we used the definition of $\sigma_K$, Eq.~\eqref{eq:kubosigma} in the main text.

Therefore, 
\begin{align}
    \left| \frac{d \, \corr_{\text{Kubo}}(t)}{dt} \right| \leq \sigma_K,
\end{align}
as claimed in the main text.

%%%%%%%%%%%%%%%
%%%%%%%%%%%%%%%

\section{Decay rates for random Hamiltonians}\label{app:random}

We focus on a model of a quantum system in which the eigenbasis of the Hamiltonian is chosen randomly as $H_U=U H U^\dagger$, 
where we average over $U$ drawn from the Haar measure on the unitary group.  We now calculate the rates analyzed in the main text, by performing analytical calculations consisting on those Haar averages, and show that for typical random Hamiltonians, the timescales coincide. These calculations are done with the Mathematica package RTNI \cite{RTNI2019}.

For a given state $\ket{\Psi}$ and observable $A$, we can write the decay rates as
\begin{align}
    &\sigma_A^2 =- \frac{\langle [H_U,[H_U,A]]\rangle}{\langle A(0) \rangle}, \\
    &\sigma_G^2 = \frac{\tr{\mathcal{D}_U(\ketbra{\Phi}{\Phi})[A,H_U]\mathcal{D}_U(\ketbra{\Phi}{\Phi})[H_U,A]}}{\tr{(\mathcal{D}_U(\ketbra{\Phi}{\Phi})A)^2}},
\end{align}
where if $H=\sum_j E_j \ket{E_j}\bra{E_j}$, the dephasing in the random eigenbasis is defined as 
\begin{equation}\label{eq:dephasedstate}
    \mathcal{D}_U(\ketbra{\Phi}{\Phi} )= \sum_j U \ket{E_j}\bra{E_j} U^\dagger \ketbra{\Phi}{\Phi} U \ket{E_j}\bra{E_j} U^\dagger.
\end{equation}
For the Kubo function, we consider the limit $\beta \rightarrow 0$, which is such that
\begin{align}
   \lim_{\beta \rightarrow 0} \frac{\corr_{\text{Kubo}}(t)}{\corr_{\text{Kubo}}(0)}= \frac{\tr{A(t)A}}{\tr{A^2}},
\end{align}
and thus in this model we have the corresponding rate
\begin{equation}
    \sigma_K^2= \frac{\tr{[H_U,A][A,H_U]}}{\tr{A^2}}.
\end{equation}
Denoting the Haar average $\int_{\text{Haar}} \cdot \, \text{d}U = \langle \cdot \rangle_U$, let us first calculate $\langle    \sigma_A^2 \rangle_U$.
\begin{align}
\langle \sigma_A^2 \rangle_U &= -\left \langle\frac{\langle [H_U,[H_U,A]]\rangle}{\langle A(0) \rangle}\right \rangle_U
\\&=-\frac{1}{\langle A(0) \rangle} \Big( \Big\langle \tr{\ketbra{\Phi}{\Phi} H_U^2 A}+ \tr{\ketbra{\Phi}{\Phi}A H_U^2 } - 2\tr{\ketbra{\Phi}{\Phi} H_U A H_U} \Big\rangle_U \Big).
\end{align}
Since the Hamiltonian appears twice, this sum of expectation values is computed with the first and second moments of the Haar measure. The result, to leading order in the inverse of the system's dimension $d^{-1}$, is
\begin{equation}\label{eq:UsigmaA}
    \langle    \sigma_A^2 \rangle_U = 2\left(\langle H^2 \rangle_{\text{MC}}-\langle H \rangle_{\text{MC}}^2  \right) \frac{(\langle A(0) \rangle-\langle A \rangle_{\text{MC}})}{\langle A(0) \rangle}  + \mathcal{O}\left(\frac{1}{d^2} \right),
\end{equation}
where $\langle A \rangle_{\text{MC}}=\frac{\tr{A}}{d}$ is the microcanonical average. Notice that our assumption of $\tr{\mathcal{D}(\ketbra{\Phi}{\Phi})A}=0$ from the main text here translates to $\langle A \rangle_{\text{MC}}=0$.

Now we calculate the other rate, which is significantly more involved. It reads 
\begin{equation}
   \langle \sigma_G^2 \rangle_U=  \left \langle     \frac{\tr{\mathcal{D}_U(\ketbra{\Phi}{\Phi})[A,H_U]\mathcal{D}_U(\ketbra{\Phi}{\Phi})[H_U,A]}}{\tr{(\mathcal{D}_U(\ketbra{\Phi}{\Phi})A)^2}} \right \rangle_U .
\end{equation}
 As a first simplification, we use the so-called ``annealed approximation", which states that we can approximate the average of the ratio is similar to the ratio of the averages
\begin{equation}\label{eq:rativoaverage}
    \langle \sigma_G^2 \rangle_U\simeq     \frac{\left \langle  \tr{\mathcal{D}_U(\ketbra{\Phi}{\Phi})[A,H_U]\mathcal{D}_U(\ketbra{\Phi}{\Phi})[H_U,A]} \right \rangle_U}{\left \langle \tr{(\mathcal{D}_U(\ketbra{\Phi}{\Phi})A)^2} \right\rangle_U}.
\end{equation}
As explained in Appendix \ref{app:annealed} below, this approximation can be made rigorous through concentration arguments and Levy's lemma. Let us now calculate the numerator and denominator separately. 
If we decompose the dephased states as in Eq. \eqref{eq:dephasedstate}, and define $U \ket{E_j}\bra{E_j} U^\dagger \equiv P^j_U$, the numerator of Eq. \eqref{eq:rativoaverage} is 
\begin{align}
\sum_{j,k}   2 \left \langle  \tr{P^j_U\ketbra{\Phi}{\Phi}P^j_U A H_U P^k_U\ketbra{\Phi}{\Phi} P^k_U H_U A} \right \rangle_U \\- \left \langle  \tr{P^j_U\ketbra{\Phi}{\Phi}P^j_U  H_U A P^k_U\ketbra{\Phi}{\Phi} P^k_U  H_U A } \right \rangle_U \\
- \left \langle  \tr{P^j_U\ketbra{\Phi}{\Phi}P^j_U  A H_U P^k_U\ketbra{\Phi}{\Phi} P^k_U A H_U  } \right \rangle_U.
\end{align}
Due to cancellations of some of the unitaries, these three correlators involve at most four pairs $\{U,U^\dagger\}$, and can thus be calculated with the fourth moment of the Haar measure. Because of this, it is an analytical expression with $3 \times 4!^2 = 1728$ terms, for which then the sum over $j,k$ has to be taken. This sum can then be simplified to
\begin{align}2  &
    \frac{\left(\langle H^2 \rangle_{\text{MC}}- \langle H \rangle_{\text{MC}}^2\right) }{(d-1) (d+1) (d+2) (d+3)} \times  \\ \Big(2 \left(d^2-1\right) \langle A(0)^2 \rangle+\langle A(0) \rangle^2 \left(d^2+d+2\right) &-2 \langle A(0) \rangle \langle A \rangle_{\text{MC}} d (3 d+1)+d \left((d+1)^2 \langle A^2 \rangle_{\text{MC}}-\langle A \rangle_{\text{MC}}^2 (d-1) d\right)\Big). \nonumber
\end{align}
The denominator on the other hand consists of a single correlator, which can be written as 
\begin{align}
    \sum_{j,k}  \tr{P^j_U\ketbra{\Phi}{\Phi}P^j_U A P^k_U\ketbra{\Phi}{\Phi} P^k_U A}.
\end{align}
This still requires the 4th moment of the Haar measure, and involves $4!^2=576$ terms. With the sums over $j,k$, they simplify to the expression
\begin{align}
   \frac{1}{d (d+1) (d+2) (d+3)} \Big(d &\left(\langle A \rangle_{\text{MC}}^2 d (d+1)+d (d+4) \langle A \rangle_{\text{MC}}^2+2 (d+4) \langle A(0)^2 \rangle+\langle A \rangle_{\text{MC}}^2\right) \\&+\langle A(0) \rangle^2 (d (d+5)+2)+2 \langle A(0) \rangle \langle A \rangle_{\text{MC}} (d-1) d+-2 \langle A(0)^2 \rangle \Big).\nonumber
\end{align}
Now, with the denominator and numerator, their ratio to leading order yields the average rate
\begin{equation}\label{eq:UsigmaG}
  \langle \sigma_G^2 \rangle_U\simeq    2 \left(\langle H^2 \rangle_{\text{MC}}- \langle H \rangle_{\text{MC}}^2\right)\frac{\langle A(0) \rangle- \langle A \rangle_{\text{MC}}}{\langle A(0) \rangle}+ \mathcal{O}\left(\frac{1}{d} \right).
\end{equation}

We end with the computation of the Kubo decay rate
\begin{align}
\langle \sigma_K^2 \rangle_U = \frac{2}{\tr{A^2}}\left(\langle \tr{H_U A H_U A-A^2 H_U^2}\rangle_U \right).
\end{align}
This again only requires the second moment of the Haar measure, from which it follows that
\begin{align}\label{eq:haarkubo}
    \langle \sigma_K^2 \rangle_U &=2d^2\frac{(\langle A^2\rangle_{\text{MC}}-\langle A\rangle_{\text{MC}}^2)(\langle H^2\rangle_{\text{MC}}-\langle H\rangle_{\text{MC}}^2)}{(d^2-1)\langle A^2\rangle_{\text{MC}}}
    \\ &=2\left( 1-\frac{\langle A\rangle_{\text{MC}}^2}{\langle A^2\rangle_{\text{MC}}}\right)\Big(\langle H^2\rangle_{\text{MC}}-\langle H\rangle_{\text{MC}}^2\Big) + \mathcal{O}\left(\frac{1}{d^2}\right).
\end{align}

We can now compare Eq. \eqref{eq:UsigmaA}, Eq. \eqref{eq:UsigmaG} and Eq. \eqref{eq:haarkubo}. We see that when we set the thermal or long-time value to zero $\langle A \rangle_{\text{MC}}=0$ (as discussed in the main text, this is necessary to compare them given the definition of the correlation functions), the three timescales coincide up to leading order
\begin{equation}
   \langle \sigma_G^2 \rangle_U \simeq \langle \sigma_A^2 \rangle_U + \mathcal{O}\left(\frac{1}{d} \right)= \langle \sigma_K^2 \rangle_U + \mathcal{O}\left(\frac{1}{d} \right) = 2 \Big(\left\langle H^2 \right\rangle_{\text{MC}}- \left\langle H \right\rangle_{\text{MC}}^2\Big)+ \mathcal{O}\left(\frac{1}{d} \right).
\end{equation}
This is the energy variance in the microcanonical distribution of the Hamiltonian $H$.

\section{The annealed approximation}\label{app:annealed}

In Eq. \eqref{eq:rativoaverage} we assumed that the average of the ratio of the correlators is approximately equal to the ratio of their averages. This has been previously referred to in the literature as the ``annealed" approximation~\cite{meir1995stochastic,PhysRevE.72.061905,cotler2017black,cotler2017chaos,ChenuQuantum2019,PhysRevLett.122.014103,PhysRevX.10.031026,shtanko2020classical}. Its general form is as follows. Given two functions two functions $f,g: U(d) \rightarrow \mathcal{R}$, such that $g(U) > 0 $, then
\begin{equation}
     \left \langle \frac{f}{g} \right\rangle_U \simeq \frac{ \langle f \rangle_U}{ \langle g \rangle_U}
\end{equation}

We now show why concentration bounds imply that this approximation is very often accurate. For the two functions $f,g$, let us write
\begin{equation}
    \langle f \rangle_U = \int_{\text{Haar}} \text{d}U f(U)= \int_{\text{Haar}} \text{d}U \frac{f(U)}{g(U)}g(U).
\end{equation}
Defining the deviation $g(U)=\langle g \rangle_U+\delta_U$, we write
\begin{align}\label{eq:annealed1}
 \langle f \rangle_U =  \int_{\text{Haar}} \text{d}U \frac{f(U)}{g(U)}g(U) = \left \langle \frac{f}{g} \right\rangle_U \langle g \rangle_U +  \int_{\text{Haar}} \text{d}U \frac{f(U)}{g(U)} \delta_U.
\end{align}
We thus need to show that the second term is small. To do so, let us define the following quantities
\begin{align}
    K_1 &\equiv \max_U \left \vert \frac{f(U)}{g(U)} \right \vert, \\
    K_2 &\equiv \frac{ \max_U g(U)}{\langle g \rangle_U}, \\
    K_3 &\equiv \min\left\{ K: \, \forall \, U,V \,\,\frac{ \vert g(U) - g(V) \vert}{\langle g \rangle_U} \le K \vert\vert U-V \vert \vert_2 \right\}.
\end{align}
The first two are the results of optimizations, and the last is the Lipschitz constant of $g(U)/\langle g \rangle_U$.  We now divide the Haar average into two and bound
\begin{align}
    \left |\int_{\text{Haar}} \text{d}U \frac{f(U)}{g(U)} \delta_U \right | &\le  \left | \int_{\vert \delta_U \vert \le \varepsilon} \text{d}U \frac{f(U)}{g(U)} \delta_U \right \vert + \left \vert\int_{\vert \delta_U \vert > \varepsilon} \text{d}U \frac{f(U)}{g(U)} \delta_U \right | \\ & \le \varepsilon K_1 + K_1 K_2  \langle g \rangle_U \times  \text{Prob}(\vert \delta_U \vert > \varepsilon).
\end{align}
The second term can be upper bounded with Levy's lemma for the Haar distribution \cite{anderson_guionnet_zeitouni_2009}, which states that
\begin{equation}
    \text{Prob}(\vert \delta_U \vert > \varepsilon) \le \text{exp}\left ( -\frac{d \varepsilon^2}{4 \langle g \rangle_U K_3} \right).
\end{equation}
This finally allows us to write, from Eq. \eqref{eq:annealed1},
\begin{equation}
    \left \langle \frac{f}{g} \right\rangle_U = \frac{ \langle f \rangle_U}{ \langle g \rangle_U} + \varepsilon',
\end{equation}
where $\vert  \varepsilon' \vert \le K_1 \left(  \frac{\varepsilon}{\langle g \rangle_U}  + K_2  \text{exp}\left ( -\frac{d \varepsilon^2}{4 K_3} \right)  \right) $. For instance, under the assumption that $K_1,K_2,K_3 \le \mathcal{O}(\text{polylog} (d))$, choosing $\varepsilon = \langle g \rangle_U d^{-1/2} \times \text{polylog} (d) $ yields the bound $\varepsilon ' \le \text{poly}(d^{-1})$. 

In our case, we have 
\begin{align}
    f(U)&= \tr{\mathcal{D}_U(\ketbra{\Phi}{\Phi})[A,H_U]\mathcal{D}_U(\ketbra{\Phi}{\Phi})[H_U,A]} \\
    g(U)&=\tr{(\mathcal{D}_U(\ketbra{\Phi}{\Phi})A)^2}.
\end{align}
The assumption that $g(U)>0$ holds here since $g(U)$ it is a trace of two positive matrices $\mathcal{D}_U(\ketbra{\Phi}{\Phi})$ and $A \mathcal{D}_U(\ketbra{\Phi}{\Phi}) A$.
To prove a more explicit bound on the error of the annealed approximation, one needs to give upper bounds on the constants $K_i$, which can in principle be obtained from the explicit expressions of $f(U)$ and $g(U)$. The assumption that $K_i \le \mathcal{O}(\text{polylog} (d))$ is likely to be satisfied in this case, since it requires that those constants grow at most polynomially in the system size.

\end{document}